
\input harvmac.tex

\def\alzi{Al. B. Zamolodchikov, `TBA Equations for Integrable
Perturbed
$SU(2)_k \otimes SU(2)_l$ $/ SU(2)_{k+l}$ Coset Models',
preprint ENS-LP-331 (1991), and references therein.}
\def\Mat{S. D. Mathur, `Quantum Kac-Moody Symmetry in Integrable
Field Theories', preprint HUTMP-90-B/299 (1990).}
\def\BMP{P. Bouwknegt, J. McCarthy and K. Pilch,
Commun. Math. Phys. 131 (1990) 125.}
\def\PasS{V. Pasquier and H. Saleur, Nucl. Phys. B330 (1990) 523.}
\def\LMC{M. L\"assig, G. Mussardo and J. L. Cardy,
Nucl. Phys. B348 (1991) 591.}

\def\reshp{N. Yu. Reshetikhin, `S-matrices in Integrable Models of
Isotropical Magnetic Chains', Harvard preprint HUTMP-90-B292 (1990);
V. V. Bazhanov and N. Yu. Reshetikhin, preprint RIMS-723 (1990). }
\def\Frohs{J. Fr\"ohlich, Commun. Math. Phys. 47 (1976) 269;
`Statistics of Fields, the Yang-Baxter
Equation, and the Theory of Knots and Links',  Cargese Lectures 1987,
G. 't Hooft et. al. (eds.) , Plenum, NY. }
\def\KMii{T. Klassen and E. Melzer, `Spectral Flow Between
Conformal Field Theories in 1+1 Dimensions',  April 1991 preprint
EFI 91-17/UMTG-162.}
\def\Fatz{V. A. Fateev,
Int. J. Mod. Phys. A6 (1991) 2109.}
\def\FW{G. Felder and C. Wieczerkowski,
Commun. Math. Phys. 138 (1991) 583. }
\def\AGT{P. C. Argyres, J. M. Grochocinski, and S.-H. H. Tye,
`Structure Constants of the Fractional Supersymmetry Chiral
Algebras', April 1991 preprint CLNS 91/1059.}
\def\ahn{C. Ahn, Nucl. Phys. B354 (1991) 57.}
\def\ntwo{P. Fendley, S. D. Mathur, C. Vafa, and N. P. Warner,
Phys. Lett. 243B (1990) 257; P. Fendley, W. Lerche, S. D. Mathur,
and N. P. Warner, Nucl. Phys. B348 (1991) 66.}

\def\Mand{S. Mandelstam, Phys. Rev. D11 (1975) 3026.}

%
%
%

\def\Nak{T. Nakatsu,
Nucl. Phys. B356 (1991) 499.}
\def\VeFa{H. J. de Vega and V. A. Fateev, `Factorizable
S-matrices for Perturbed W-invariant Theories,'
preprint LPTHE 90-36, July 1990. }

\def\Kor{V. E. Korepin, Teor. Mat. Fiz. 41 (1979) 169.}

\def\Geri{J.-L. Gervais,
Commun. Math. Phys. 130 (1990) 257.}

\def\Smirii{F. A. Smirnov, Commun. Math. Phys. 132 (1990) 415.}
\def\Smiriii{F. A. Smirnov,
Int. J. Mod. Phys. A6 (1991) 1253.}

\def\Hollow{T. Hollowood, `A Quantum Group Approach to Constructing
Factorizable S-Matrices,' Oxford Preprint OUTP-90-15P, June 1990. }

\def\Yang{D. Bernard,
Commun. Math. Phys. 137 (1991) 191. }

\def\fssg{D. Bernard and A. LeClair,
 Phys. Lett. B247 (1990) 309.}
\def\ABL{C. Ahn, D. Bernard, and A. LeClair,
Nucl. Phys. B346 (1990) 409. }

\def\cosetFF{D. Kastor, E. Martinec, and Z. Qiu, Phys. Lett. B 200
(1988) 434\semi
J. Bagger, D. Nemeschansky, and S. Yankielowicz, Phys. Rev. Lett. 60
(1988) 389\semi
F. Ravanini, Mod. Phys. Lett. A 3 (1988) 397\semi
\chrav }


\def\cosetFF{D. Kastor, E. Martinec, and Z. Qiu, Phys. Lett. B 200
(1988) 434\semi
J. Bagger, D. Nemeschansky, and S. Yankielowicz, Phys. Rev. Lett. 60
(1988) 389\semi
F. Ravanini, Mod. Phys. Lett. A 3 (1988) 397\semi
\chrav }


\def\chrav{P. Christe and F. Ravanini, Int. Jour. Mod. Phys. 4 (1989) 897.}

\def\Zflow{A. B. Zamolodchikov, Sov. J. Nucl. Phys. 46 (1987) 1090.}

\def\BLii{D. Bernard and A. LeClair,
Nucl. Phys. B340 (1990) 721.}

\def\rsg{A. LeClair, Phys. Lett. 230B (1989) 103;
Cornell preprint CLNS 89/955,
in  proceedings of the 1989 Banff workshop on `Physics, Geometry,
and Topology',
Plenum Press,  H. C. Lee, ed. }
\def\RS{N. Yu. Reshetikhin and F. Smirnov,
Commun. Math. Phys. 131 (1990) 157. }

\def\Zamovi{A. B. Zamolodchikov, Sov. J. Nucl. Phys. 44(1986) 529. }

\def\Dotsenko{V. Dotsenko and V. Fateev, Nucl. Phys. B240 (1984) 312.}
\def\BPZ{A. A. Belavin, A. M. Polyakov, and A. B. Zamolodchikov,
Nucl. Phys. B241 (1984) 333.}

\def\Drinfeld{V. G. Drinfel'd, Sov. Math. Dokl. 32 (1985) 254;
Sov. Math. Dokl. 36 (1988) 212. }
\def\Jimbo{M. Jimbo, Lett. Math. Phys. 10 (1985) 63; Lett. Math. Phys. 11
(1986) 247; Commun. Math. Phys. 102 (1986) 537.}

\def\FQS{D. Friedan, Z. Qiu, and S. Shenker, Phys. Rev. Lett. 52 (1984)
1575.}

\def\ABF{G. Andrews, R. Baxter, and P. J. Forrester, Jour. Stat.
Phys. 35 (1984) 193.}
\def\Kyoto{E. Date, M. Jimbo, A. Kuniba, T. Miwa, and M. Okado,
Nucl. Phys. B 290 (1987) 231.}

\def\Pasquierii{V. Pasquier, Commun. Math. Phys. 118 (1988) 355. }
\def\STF{E. K. Sklyanin, L. A. Takhtadzhyan, and L. D. Faddeev, Theor.
Math. 40 (1980) 688.}

\def\Zamolodchikovi{A. B. Zamolodchikov and Al. B. Zamolodchikov, Annals
Phys. 120 (1979) 253.}
\def\Colemani{S. Coleman, Phys. Rev. D 11 (1975) 2088.}

\def\Kirillovi{A. N. Kirillov and N. Reshetikhin, LOMI preprint E-9-88,
in  proceedings of Marseille conference, V. Kac, ed.,
World Scientific . }

\def\TK{A. Tsuchiya and Y. Kanie, Lett. Math. Phys. B (1987) 303.}
\def\MSi{G. Moore and N. Seiberg, Phys. Lett. B 212 (1988) 451;
Commun. Math. Phys. 123(1989) 177.}

\def\GoS{C. Gomez and G. Sierra, Phys. Lett. B240 (1990) 149;
Nucl. Phys. B352 (1991) 791.}
\def\DBqad{D. Bernard, Lett. Math. Phys. 17 (1989) 239.}

\def\FFK{G. Felder, J. Fr{\"{o}}hlich and G. Keller,
Comm. Math. Phys. 124 (1989) 647.}
\def\Felderi{G. Felder, Nucl. Phys. B317 (1989), 215. }
\def\MR{G. Moore and N. Reshetikhin,
Nucl. Phys. B328 (1989) 557.}
\def\CThorn{T. L. Curtright and C. B. Thorn, Phys. Rev. Lett. 48(1982)
1309.}

\def\EYall{T. Eguchi and S.-K. Yang,
Phys. Lett. B224 (1989) 373; ibid. 235 (1990) 282.}

\def\Zamoiii{A. B. Zamolodchikov, Int. Journ. of Mod. Phys. A4 (1989) 4235;
in Adv. Studies in Pure Math. vol. 19 (1989) 641.}
\def\Smiri{F. A. Smirnov, Int. Jour. Mod. Phys. A4 (1989) 4213; Nucl. Phys.
B337 (1990) 156.}

\def\FRS{K. Fredenhagen, K.H. Rehren, and B. Schroer,
Commun. Math. Phys. 125: 201, (1989).}
%

%


%
%
%

%
%

\def\bra#1{{\langle #1 |  }}
\def\lb{ \left[ }
\def\rb{ \right]  }
\def\tilde{\widetilde}
\def\bar{\overline}
\def\hat{\widehat}
\def\*{\star}
\def\[{\left[}
\def\]{\right]}
\def\({\left(}		\def\BL{\Bigr(}
\def\){\right)}		\def\BR{\Bigr)}
	\def\BBL{\lb}
	\def\BBR{\rb}
%
%
\def\zb{{\bar{z} }}
\def\frac#1#2{{#1 \over #2}}
\def\inv#1{{1 \over #1}}
\def\half{{1 \over 2}}
\def\d{\partial}

\def\vev#1{\langle #1 \rangle}
\def\ket#1{ | #1 \rangle}

\def\2pi{\hbox{$2\pi i$}}

\def\dsl{\raise.15ex\hbox{/}\kern-.57em\partial}
\def\Dsl{\,\raise.15ex\hbox{/}\mkern-.13.5mu D}
%
%
\def\th{\theta}		\def\Th{\Theta}
\def\ga{\gamma}		
\def\be{\beta}
\def\al{\alpha}
\def\ep{\epsilon}
\def\la{\lambda}	
		\def\De{\Delta}

\def\vphi{\varphi}
%
%
\def\CA{{\cal A}}		\def\CC{{\cal C}}
		\def\CF{{\cal F}}
	\def\CH{{\cal H}}	
\def\CJ{{\cal J}}		
		\def\CO{{\cal O}}
		
		\def\CU{{\cal U}}

\def\2pi{\hbox{$2\pi i$}}

\def\dsl{\raise.15ex\hbox{/}\kern-.57em\partial}
\def\Dsl{\,\raise.15ex\hbox{/}\mkern-.13.5mu D}
%
%
%
\font\numbers=cmss12
\font\upright=cmu10 scaled\magstep1
\def\stroke{\vrule height8pt width0.4pt depth-0.1pt}
\def\topfleck{\vrule height8pt width0.5pt depth-5.9pt}
\def\botfleck{\vrule height2pt width0.5pt depth0.1pt}
\def\Zmath{\vcenter{\hbox{\numbers\rlap{\rlap{Z}\kern 0.8pt\topfleck}\kern
2.2pt
                   \rlap Z\kern 6pt\botfleck\kern 1pt}}}
\def\Qmath{\vcenter{\hbox{\upright\rlap{\rlap{Q}\kern
                   3.8pt\stroke}\phantom{Q}}}}
\def\Nmath{\vcenter{\hbox{\upright\rlap{I}\kern 1.7pt N}}}
\def\Cmath{\vcenter{\hbox{\upright\rlap{\rlap{C}\kern
                   3.8pt\stroke}\phantom{C}}}}
\def\Rmath{\vcenter{\hbox{\upright\rlap{I}\kern 1.7pt R}}}
\def\Z{\ifmmode\Zmath\else$\Zmath$\fi}
\def\Q{\ifmmode\Qmath\else$\Qmath$\fi}
\def\N{\ifmmode\Nmath\else$\Nmath$\fi}
\def\C{\ifmmode\Cmath\else$\Cmath$\fi}
\def\R{\ifmmode\Rmath\else$\Rmath$\fi}

 \Title{CLNS 91/1099}
 {\vbox{\centerline{Restricted Quantum Affine Symmetry   }
 \centerline{ of Perturbed Minimal Conformal Models} }}

 \bigskip
 \bigskip
 \centerline{G. FELDER}
 \centerline{Mathematik}
 \centerline{ETH-Zentrum}
 \centerline{8092 Z\"urich, Switzerland}
 \bigskip
 \smallskip

 \centerline{A. LECLAIR}
 \centerline{Newman Laboratory}
 \centerline{Cornell University}
 \centerline{Ithaca, NY  14853, USA}

 \vskip .3in
\def\slqh{{ $\CU_q \( \hat{sl(2)}  \)$ }}

 We study the structure of superselection sectors of
 an arbitrary perturbation of a conformal field theory.
 We describe how a restriction of the q-deformed $\hat{sl(2)}$ affine
 Lie algebra symmetry of the sine-Gordon theory can be used to derive
 the S-matrices of the $\Phi^{(1,3)}$ perturbations of the minimal
 unitary series.  This analysis provides an identification of fields
 which create the massive kink spectrum. We investigate the ultraviolet limit
 of the restricted sine-Gordon model, and explain the relation between the
 restriction and the Fock space cohomology of minimal models.
 We also comment on the structure of degenerate vacuum states.
 Deformed Serre relations are proven for
 arbitrary affine Toda theories, and it is shown in certain cases how
 relations of the Serre type
 become fractional spin supersymmetry relations upon
 restriction.

 \vskip 10pt


\Date{8/91}
%
%

%
%
%
%
%
%

%
%
%
%

\def\jua{{ J^a_\mu }}
\def\FrKe{J. Fr\"ohlich and T. Kerler,
`On the role of Quantum Groups in Low Dimensional Quantum Field Theory,'
 to appear.}
\def\BLnlc{D. Bernard and A. LeClair,
 `Quantum Group Symmetries and Non-Local Currents in 2D QFT,'
to appear in
Commun. Math. Phys.   }
\def\prin{ {\rm prin.} }
\def\TH{\Theta}
\def\THb{{\bar{\Theta}}}
\def\Th{\TH}
\def\Thb{\THb}
\def\Psib{{ \bar{\Psi} }}
\def\ad#1{{ {\rm ad}_{#1} }}
\def\vphi{\varphi}
\def\slh{{ $\hat{sl(2)}$ }}
\def\slq{{ $\CU_q \( sl(2) \)$ }}
\def\slqh{{ $\CU_q \( \hat{sl(2)}  \)$ }}

\def\phib{{ \bar{\phi} }}
\def\Ph#1#2{{ \Phi^{(#1,#2)} }}
\def\CHb{ {\bar{\CH}} }
\def\ph#1#2{{ \phi^{(#1,#2)} }}

\def\phb#1#2{{ \phib^{(#1,#2)} }}
\def\Qb{{ \bar{Q} }}
\def\bQ{{ \bar{Q} }}
\def\Jb{{ \bar{J} }}
\def\a{{\rm ad}}

\def\CHb{{ \bar{\CH} }}
\def\UG{{ \CU_q \( \hat{G}  \) }}
\def\UGm{{ $\CU_q \( \hat{G}  \) $}}
\def\Gh{{\hat{G}}}
\def\vphib{{\bar{\varphi} }}
\font\eightrm=cmr8

\newsec{Introduction}

The domain of integrable massive quantum field theory in
1+1 dimensions has broadened considerably over the past few
years to include many new models.  This is due primarily
to the idea that massive theories can be formulated as
integrable perturbations of conformal field theories (CFT),
as originally put forward by A. Zamolodchikov\ref\rzamoiii{\Zamoiii}.
Consequently, if one perturbs relatively novel conformal
field theories such as the minimal conformal series
\ref\rBPZ{\BPZ}\ref\rFQS{\FQS}, one is certain to obtain
new and interesting massive theories.  Exact S-matrices have
now been proposed for certain perturbations of many infinite
series of unitary and non-unitary conformal field theories.
Some unitary examples include the $\Ph 13$ perturbations
\ref\rsss{\Smiri}\ref\rey{\EYall}
\ref\rrsg{\rsg}
\ref\rRS{\RS}
\ref\rBLii{\BLii}\
and $\Ph 12 , ~\Ph 21$ perturbations \ref\rsmiriii{\Smiriii}\
of
the $c<1$ minimal unitary models, W-invariant theories \ref\rABL{\ABL}
\ref\rNak{\Nak}
\ref\rHollow{\Hollow}\ref\rVeFa{\VeFa},
the cosets $G_k \otimes G_l /G_{k+l}$\rABL ,
$Z_k$ parafermions \ref\rFat{\Fatz},
and $Z_k$ parafermions
coupled to a single boson (fractional supersymmetric sine-Gordon theories),
which includes the special cases of
N=2 superconformal series and level-k Wess-Zumino-Witten
theories \ref\rfssg{\fssg} \ref\rahn{\ahn} .
Perturbations of N=2 superconformal field theories have also been
studied in a Landau-Ginzburg approach \ref\rWar{\ntwo}.
Many of these models are quantum field theory versions of solvable
models of lattice statistical mechanics \ref\rABF{\ABF}\ref\rKyoto{\Kyoto}.
Indeed some of the above S-matrices were obtained in \ref\rbethe{\reshp}\
from a lattice Bethe-ansatz approach, which is close in
spirit to the statistical mechanics methodology.

In this paper we will be concerned with the $\Ph 13$ perturbation of
the minimal unitary series, with central charge
\eqn\Ii{
c= 1- \frac{6}{p(p+1)} ~~~~~~p=3,4,...}
This model is formally described by the Euclidean action
\eqn\Iib{
S\ =\ S_{{\rm CFT}} + \frac{\la}{2\pi }
	\int d^2z\ \Ph 13 (z,\zb) .}
The conjectured properties of these models are as follows.
 The massive spectrum consists of kinks $K_{jk}$,
$j,k \in \{ 0, 1/2 , 1, ..., p/2 -1 \}$, $j=k\pm 1/2$.
The scattering of these kinks is described by an exact S-matrix
of the so-called RSOS form, which can be derived by a
quantum group restriction of the conjectured S-matrix for
sine-Gordon solitons.  These kinks have the appealing
physical interpretation as connecting  $p-1$
degenerate vacua in a Landau-Ginzburg description of the conformal
model \ref\rz{A. Zamolodchikov, Landau Institute preprint,
September 1989, unpublished.}\rBLii .  These models, which have
come to be known as the restricted sine-Gordon theories (RSG), are
further characterized by some fractional spin supersymmetries
\rz\rBLii .  Recent  support for these conjectures has been
provided in studies of the thermodynamic Bethe ansatz
\ref\rAZam{\alzi} \ref\rKMii{\KMii}.

Though the above properties of the RSG theories are generally
accepted,
 a complete quantum field theoretic basis for some
of the essentially on-shell arguments was missing in the original
works.  This was due in part to the absence of a satisfactory
non-perturbative
derivation of the usual (unrestricted) sine-Gordon soliton S-matrix
\ref\rZamoi{\Zamolodchikovi}\foot{The standard non-perturbative framework
for quantizing the SG theory is the quantum inverse scattering method
(QISM) \ref\rSTF{\STF} , which is an algebraization of Bethe-ansatz
methods. Unfortunately the soliton sectors are not easily dealt with
in this framework.  For a Bethe-ansatz derivation of the SG soliton
S-matrix in the Thirring description, see \ref\rKor{\Kor}.}.
More recently in \rfssg\ref\nlc{\BLnlc}\
a derivation of this latter S-matrix was given based on the construction
of some non-local conserved currents in the sine-Gordon (SG) model that
generate the q-deformation of $sl(2)$ affine Lie algebra
 \slqh \ref\rJimbo{\Jimbo}\ref\rDrin{\Drinfeld} .
Quantum affine algebras in connection with $\Ph 13$ perturbed minimal
CFT was
also discussed in \ref\rMathur{\Mat} \foot{However there are
significant differences in the proposals of \nlc\ and \rMathur\
that are not easily reconciled.  The differences are apparently
due to the fact that in \nlc\ and the present work, one is
only concerned with true symmetries of the theory that are
generated by quantum conserved charges which commute with the Hamiltonian and
S-matrix.  On the other hand the quantum affine structure in
\rMathur\ is more in the spirit of the quantum group symmetry
of CFT, which encodes certain aspects of the fusion rules, and
is  not a symmetry in the above sense.}.

The purpose of this paper is to provide a quantum field theoretic
derivation of the RSG S-matrices based on an analysis of their
non-local conserved currents.  These quantum symmetries are
inherited from the  \slqh symmetry of the SG model, and are related
to the fractional supersymmetries described in \rBLii , in a way we
will make precise.  However the algebra satisfied by the conserved
charges is no longer \slqh , but rather a `restriction' of it.
Our treatment identifies the fields
which create the quantum kinks as the intertwiners for the
{\it chiral} fields $\ph 21$  or $\phb 21$.  Based on the
study of the thermodynamic Bethe ansatz equations of Al. Zamolodchikov
\rAZam , Klassen and Melzer \rKMii\ conjectured that the degenerate vacua
were associated with the local CFT states $\Ph nn (0)\ket{0}$.
We offer a justification of this identification within our framework.

The general theory of superselection sectors, as outlined in
\ref\rFroh{\Frohs}\ref\rFRS{\FRS},
provides the proper conceptual understanding of some aspects of
the RSG quantum field theory.  We provide a simple characterization
of the superselection sectors of an arbitrary perturbed CFT in the
sequel.  To our knowledge the RSG theories represent the first example
of a model of massive particles with non-abelian sectors.

In the next section we will complete the identification of the
\slqh symmetry in the SG theory by proving that the conserved charges
satisfy the appropriate deformed Serre relations.
We also extend this analysis to the generalization of SG theory
to an arbitrary affine Toda theory.  This computation
is similar but not identical to the  computations in
\ref\rBMP{\BMP}\rMathur ; rather
our computation is closer to the formulation of twisted homology
\ref\rFW{\FW}.
Sections 3 and 4 contain the main results outlined above.

\newsec{Quantum Affine Symmetry of the Sine-Gordon and Affine
Toda Theories}
\def\Gh{{\hat{G}} }
\def\av{{\vec{\al}}}
\def\vphi{\varphi}
\def\vphib{{\bar{\varphi} }}

\bigskip
\noindent
2a.  {\it Quantum Affine Algebras}
\medskip
We review a few basis facts about the
quantum affine algebras\rJimbo\rDrin .
Let $\av_i$, $i=1,..,{\rm rank}(\hat{G} )$ denote a basis of simple roots
of an arbitrary, possibly twisted,  affine Lie algebra $\hat{G}$, and
$a_{ij} = 2\av_i \cdot \av_j / |\av_i |^2 $ its generalized
Cartan matrix.  Untwisted $\hat{G}$
are obtained from a simple Lie algebra $G$ by appending the maximal root
$\av_0$ of $G$ to the simple roots of $G$, such that
${\rm rank}(\Gh ) = {\rm rank}(G) +1$.  The quantum affine algebra
$\CU_q \( \Gh \)$ is a deformation of the universal enveloping algebra
of $\Gh$ generated by $H_i , E^{\pm}_i $, $i=1,..,{\rm rank}(\Gh )$
satisfying the relations
\eqnn\IIi
$$\eqalignno{
\[ H_i , H_j \] &=0\cr
\[ H_i , E^\pm_j \] &= \pm \av_i \cdot \av_j \, E^\pm_j  &\IIi\cr
\[ E^+_i , E^-_j \] &= \delta_{ij} \>
\frac{q^{H_i} - q^{-H_i } } {q_i - q_i^{-1}} \cr}$$
where
$q_i \equiv q^{ |\av_i |^2 /2}$, and $q$ is a free parameter.
The complete set of relations includes the
additional deformed Serre relations
\eqn\IIii{
\sum_{\nu=0}^{1-a_{ij}} (-)^\nu
\left[ \matrix{1-a_{ij}\cr \nu\cr } \right]_{q_i}
\( E^\pm_i \)^{1-a_{ij} -\nu } E^\pm_j \( E^\pm_i \)^\nu = 0 }
where
$$\left[ \matrix{m\cr n\cr } \right]_q = \frac{[m]_q !}{[n]_q ! [m-n]_q !}
,~~~~~[m]_q ! = \prod_{1\leq i \leq m } [i]_q , ~~~~~~
[i]_q = \frac{q^i - q^{-i} }{q-q^{-1}} . $$

\def\De{{\Delta}}
\def\ot{{\otimes}}
\def\ep{\varepsilon}

The algebra $\UG$ is a Hopf algebra equipped with
comultiplication $\Delta : \UG \to \UG \ot \UG$,
counit $\ep : \UG \to \C$, and antipode $s: \UG \to \UG$,
with the following properties:
\eqna\IIiii
$$\eqalignno{
\De (a) \De (b) &= \De (ab) &\IIiii{a}\cr
\( \De \ot id \) \De (a) &= \( id \ot \De \) \De (a) &\IIiii{b} \cr
\( \ep \ot id \) \De (a) &= \( id \ot \ep \) \De (a) = a &\IIiii{c} \cr
m \( s\ot id \) \De (a) &= m \( id\ot s \) \De (a) = \ep (a) &\IIiii{d} \cr
}$$
for $a,b\in \UG$ and $m$ the multiplication map: $m(a\ot b) = ab$.
Eq. \IIiii{a}\ implies $\De$ is a homomorphism of $\UG$ to $\UG \ot \UG$,
\IIiii{b}\ is the coassociativity, and \IIiii{c,d}\ are the defining
properties of the counit and antipode.  These operations have the
following specific form
\eqnn\IIiv
$$\eqalignno{
\De (H_i ) &= H_i \ot 1 + 1\ot H_i \cr
\De (E_i^\pm ) &= E^\pm_i \ot q^{-H_i /2} + q^{H_i /2} \ot E^\pm_i
&\IIiv\cr
s(E^\pm_i ) &= -q_i^{\mp 1} E_i^\pm , ~~~~~s(H_i ) = -H_i \cr
\ep (E^\pm_i ) &= \ep (H_i ) = 0 .\cr}$$

Let us define an adjoint action $\ad a : \UG \to \UG$ for
$a\in \UG$ as follows.  If
\eqn\IIv{
\De (a) = \sum_i a_i \ot a^i }
then
\eqn\IIvi{
\ad a (b) = \sum_i a_i \, b \, s(a^i ) . }
(Above $a_i , a^i , b \in \CU_q \( \Gh \)$.)
This adjoint action is an action of $\UG$ on itself, i.e. if
$m^c_{ab} $ are the `structure constants' of $\UG$, then
\eqn\IIvii{
e_ae_b = m^c_{ab} \, e_c ~~\Rightarrow \ad {e_a}
  \ad {e_b}  = m_{ab}^c
\, \ad {e_c} . }
(See
\ref\rLS{A. LeClair and F. Smirnov, `Infinite Quantum Group Symmetry
of Fields in Massive 2D Quantum Field Theory', Cornell preprint
CLNS 91-1056, to appear in Int. Journ. Mod. Phys. A. }
\ref\rBF{D. Bernard and G. Felder, `Quantum Group Symmetries in 2D
Lattice Quantum Field Theory', Preprint SPhT-91-002, ETH-TH/91-10,
to appear in Nucl. Phys. B.}
.)
For ordinary Lie algebras, ${\rm ad}$ is just the commutator.

For reasons that will become apparent, let us define a new basis
in $\UG$ generated by $H_i , Q_i , \Qb_i $:
\eqn\IIviii{
Q_i \equiv E^+_i q^{H_i /2} , ~~~~~\Qb_i \equiv E^-_i q^{H_i /2} . }
They have the following properties as a consequence of \IIiv :
\eqnn\IIix
$$\eqalignno{
\De (Q_i ) &= Q_i \ot 1 + q^{H_i } \ot Q_i \cr
\De (\Qb_i ) &= \Qb_i \ot 1 + q^{H_i } \ot \Qb_i  &\IIix\cr
s(Q_i ) &= - q^{-H_i } Q_i , ~~~~~s(\Qb_i ) = -q^{-H_i} \Qb_i \cr
\ep(Q_i ) &= \ep (\Qb_i ) = 0 . \cr  }$$
The relations \IIi\ can now be written as
\eqnn\IIx
$$\eqalignno{
\[ H_i , Q_j \] &= \av_i \cdot \av_j \> Q_j ~~~~~~~
\[ H_i , \Qb_j \] = -\av_i \cdot \av_j \> \Qb_j \cr
\a_{Q_i } (\Qb_i ) &= Q_i \Qb_i - q_i^{-2} \Qb_i Q_i
= \frac{1-q^{2H_i}}{ q_i^{-2} -1 } . &\IIx\cr}$$
It can be shown by explicit computation that the deformed Serre
relations can be expressed in the remarkably simple form
\eqn\IIxi{
{\rm ad}_{Q_i}^{1-a_{ij}}\( Q_j \) =
{\rm ad}_{\Qb_i}^{1-a_{ij}}\( \Qb_j \) = 0.}
The above result is a variation of the presentation defined in
\ref\rDB{\DBqad} .

For \slqh the simple roots may be chosen such that
$\av_0 = -\av_1$, $|\av_1 |^2 = 2$.  The central element
$k=H_0 + H_1$
of the affine Lie algebra $\hat{sl(2)}$
is called the level.  The \slqh loop algebra is
obtained by setting $k=0$.

\bigskip
\noindent
2b. {\it The \slqh Conserved Currents of the Sine-Gordon Theory}
\medskip
\def\bh{\hat{\beta}}
\def\hb{\bh}

We now review how the \slqh loop algebra symmetry is realized
in the sine-Gordon theory\nlc .  The SG theory may be treated
as a massive perturbation of the $c=1$ conformal field theory
corresponding to a single real scalar field\foot{$z,\zb$ are
Euclidean light-cone coordinates: $z=(t+ix)/2 , \zb = (t-ix)/2$. }:
\eqn\IIxii{ S\ =\ \inv{4\pi}\int d^2z\ \d_z\Phi \d_{\zb}\Phi \
+ \frac{\la}{\pi }\ \int d^2z\ :\cos\(\hb \Phi\): ~~.}
With the above normalization of the kinetic term, the free fermion
and Kosterlitz-Thouless points occur at $\bh = 1, \sqrt{2}$
respectively\foot{$\bh$ is related to the conventional coupling
$\be$ in \ref\rCole{\Colemani}\ref\rmand{\Mand}\ by
$\bh = \be / \sqrt{4\pi}$.}.
The theory has a well-known conserved topological current
\eqn\IIxiiib{
\CJ_\mu^{\rm top.} (x) = \frac{\bh}{2\pi} \ep_{\mu\nu} \d_\nu \Phi (x) }
which defines the topological charge
\eqn\IIxiii{
T= \frac{\bh}{2\pi} \int_{-\infty}^{\infty} dx \> \d_x \Phi (x). }
With
the above $\bh$-dependent normalization of the topological charge,
the soliton states are known to have $T=\pm 1$ \rCole
\rmand .  We will give a different derivation of this
normalization in the sequel where we describe superselection sectors
of perturbed CFT.
For  applications to perturbations of the minimal unitary CFT's,
we need only consider the theory in the range
$1\leq \bh \leq \sqrt{2}$, where the only particles in the spectrum
are solitons.

Define the quasi-chiral components $\vphi , \vphib$ of $\Phi$ as
\eqn\IIxiv{\eqalign{
\vphi(x,t)\ &=\ \half\BL\Phi(x,t)
+\int_{-\infty}^x dy\ \d_t\Phi(y,t)\BR \cr
\vphib(x,t)\ &=\ \half\BL\Phi(x,t)
-\int_{-\infty}^x dy\ \d_t\Phi(y,t)\BR ,\cr}}
such that $\Phi = \vphi + \vphib$.  When $\lambda =0$,
$\vphi = \vphi (z)$, and
$$< \vphi (z) \vphi (w) > = -\log (z-w).$$
Similarly for $\vphib$.
We will make use of the following braiding relations:
\eqna\IIxivb   $$\eqalignno{
\exp\(ia\vphi(x,t)\)\ \exp\(ib\vphi(y,t)\)\ &=\
e^{\pm i\pi ab}\ \exp\(ib\vphi(y,t)\)\ \exp\(ia\vphi(x,t)\)
\ ;\ {\rm for}\ x{> \atop <}y \cr
{}~&~ &\IIxivb a\cr
\exp\(ia\vphib(x,t)\)\ \exp\(ib\vphib(y,t)\)\ &=\
e^{\mp i\pi ab}\ \exp\(ib\vphib(y,t)\)\ \exp\(ia\vphib(x,t)\)
\ ;\ {\rm for}\ x{> \atop <}y \cr
{}~&~ &\IIxivb b\cr
\exp\(ia\vphi(x,t)\)\ \exp\(ib\vphib(y,t)\)\ &=\
e^{ i\pi ab}\ \exp\(ib\vphib(y,t)\)\ \exp\(ia\vphi(x,t)\)
\ ;\ \forall\ x, y ,  \cr
{}~&~ &\IIxivb c\cr}$$
which are derived using
$$[\Phi (x,t), \d_t \Phi (y,t) ] = 4\pi i \delta (x-y).$$
The topological charge of these fields follows from the relation
\eqn\IIxivc{
\[ T, \exp \( ia\vphi + ib\vphib \) \] = \bh (a-b)
\exp \( ia \vphi + ib\vphib \) .}
The conformal dimensions $(h,\bar{h} )$ of the field
$\exp \( ia\vphi + ib\vphib \)$ are
$(a^2 /2 , b^2 /2 )$, and its Lorentz spin (eigenvalue under
Lorentz boosts) is $h-\bar{h}$.

It can be shown that the model \IIxii\ has the following non-local
quantum conserved currents:
\eqn\IIxv{\d_{\zb}J_{\pm}\ =\ \d_zH_{\pm}
\qquad ; \qquad
\d_z{\bar J}_{\pm}\ =\ \d_{\zb}{\bar H}_{\pm} ,}
where
\eqna\IIxvi
$$\eqalignno{
J_{\pm}(x,t)&=\exp\(\pm\frac{2i}{\hb}\ \vphi(x,t)\), ~~~~~
{\bar J}_{\pm}(x,t)=\exp\(\mp\frac{2i}{\hb}\ \vphib(x,t)\)
	 &\IIxvi a\cr
H_{\pm}(x,t)&=\la
\frac{\bh^2}{\bh^2 -2}
\exp\lb \pm i\(\frac{2}{\hb}-\hb\)\vphi(x,t)
	\mp i\hb\ \vphib(x,t)\rb
&\IIxvi b\cr
{\bar H}_{\pm}(x,t)&=\la
\frac{\bh^2}{\bh^2 -2}
\exp\lb \mp i\(\frac{2}{\hb}-\hb\)\vphib(x,t)
	\pm i\hb\ \vphi(x,t)\rb .
&\IIxvi c\cr
  }$$
These currents define the following conserved charges
\eqn\IIxvii{\eqalign{
Q_{\pm}\ &=\ \inv{2\pi i} \( \int dz J_{\pm} + \int d\zb H_{\pm} \) \cr
{\bar Q}_{\pm}\ &=\ \inv{2\pi i} \( \int d\zb {\bar J}_{\pm}  +
\int dz
  {\bar H}_{\pm} \) .  \cr}}

The conserved charges have non-trivial Lorentz spin:
\eqn\IIxviii{\inv{\ga}\ \equiv\ {\rm spin}\(Q_{\pm}\)\ =\
-{\rm spin}\({\bar Q}_{\pm}\)\ =\
\frac{2}{\hb^2}-1 .}
Consequently the non-local conserved currents have non-trivial
braiding relations among themselves and with other fields.
In applications to quantum field theory, many of the essential
properties of quantum groups find their origin in these
braiding relations \ref\ryang{\Yang}\nlc .  This fact was also recognized by
Gomez and Sierra in their study of quantum group symmetry
in the minimal conformal models\ref\rGoS{\GoS}.
Consider more generally some conserved currents $
\d_\mu J^a_\mu = 0$, defining some conserved charges
$Q^a = \inv{2\pi i} \int dx J^a_0 (x,t)$, and let us suppose
the following braiding relations with a set of fields $\Phi^i (x)$:
\eqn\IIxix{ J^a_\mu (x,t)\ \Phi^k(y,t) \ =\ \sum_{b,l}
 R^{ak}_{bl}\ \Phi^l(y,t)\ J^b_{\mu}(x,t) \qquad ;\qquad {\rm for} ~x<y . }
(The above ansatz for the braiding relations is not completely
general.  For certain kinds of
non-abelian braiding, the appropriate generalization
will appear in the sequel.)
The action of the charges $Q^a$ on the fields, which we will denote
as $\ad {Q^a}$, can be defined as follows
\eqn\IIxx{ {\ad {Q^a} }\BL \Phi^k(y) \BR\ =\
\inv{2\pi i} \int_{\ga(y)}dz_{\nu}\ \ep^{\nu\mu}\ \jua(z)\ \Phi^k(y) ~,}
where the contour $\ga (y)$ is drawn in figure 1.

\vskip 2cm
\ \ \ \ \ \ \ \ \ \ \ \ \ \ \ \ \ \ \ \ \ \ \ \ \ \
\ \ \ \ \ \ \ \ \ \ \ \ \ \ \ \ \ \ \ \ \ \ \ \ \ \
\ \ \  $\cdot\ y$
\vskip 2cm
\centerline{\eightrm Fig. 1. The contour of integration defining
the adjoint action in
\IIxx .}
\vskip 10pt

\noindent
Using the
braiding relations \IIxix\ one derives
\eqn\IIxxi{ {\ad{Q^a}}\BL \Phi^k(y)\BR\ =\
Q^a\ \Phi^k(y) - R^{ak}_{bl}\ \Phi^l(y)\ Q^b .}
If $R^{ab}_{dc}$ is the braiding matrix of the currents with
themselves as in \IIxix , then the integrated version of \IIxxi\ is
\eqn\IIxxii{  \ad {Q^a} \( Q^b \) =
Q^a\ Q^b - R^{ab}_{dc}\ Q^c\ Q^d  . }

The adjoint action of $Q^a$ on a product of two fields at
different spacial locations is again defined as in \IIxx ,
where now the contour surrounds the locations of both fields.
Using the braiding relations to pass the current through the
first field before acting on  the second, one finds that this
action has a non-trivial comultiplication
\eqn\IIxxiii{
\De \(  {Q^a} \) =  {Q^a} \ot 1 + \Theta^a_b \ot  {Q^b},}
where $\Theta^a_b$ is the braiding operator which acts on the
vector space spanned by the fields $\Phi^i$, i.e. $\Th^a_b$
has the matrix elements $\Th^{ak}_{bl} = R^{ak}_{bl}$.

Returning now to the SG theory, and using the braiding relations
\eqn\IIxxiv{
J_{\pm}(x,t)\ {\bar J}_{\mp}(y,t)\ =\
q^{-2}\ {\bar J}_{\mp}(y,t)\ J_{\pm}(x,t) \qquad ;\ \forall\ x,y }
where
\eqn\eqd{ q=\exp(-2\pi i /\hb^2)=-\exp(-i\pi/\ga),}
one can show that $Q_\pm , \Qb_\pm$ and $T$ together generate
the \slqh relations \IIx . The isomorphism is
\eqnn\IIxxvi
$$\eqalignno{
Q_+ &= c\> Q_1 ~~~~~;~~~~~ Q_- = c \> Q_0 \cr
\Qb_- &= c \> \Qb_1 ~~~~~;~~~~~\Qb_+ = c \> \Qb_0 &\IIxxvi\cr
T&=H_1 = -H_0 , \cr}$$
where $c$ is a constant
( $c=\frac{\la}{2\pi i} \( \frac{\bh^2}{\bh^2 -1} \)^2 (q^{-2} -1 )$ ).
 The last relation implies the
level is zero and we are actually dealing with the loop algebra.
We emphasize that the algebraic relations \IIx\ for the
conserved charges were derived in \nlc\ using the quantum
field theory expressions \IIxx\IIxxii ;  the RHS of \IIx\
was found in closing the contour in \IIxx .  Thus it is important to
notice that for the particular conserved charges we are considering,
the adjoint action defined in the quantum field theory \IIxx\ and
the formal adjoint action \IIvi\ are equivalent.  We will have
more to say about this below.

The asymptotic soliton states of topological charge $\pm 1$
are denoted $\ket{\pm 1/2,\th}$, where $\th$ is the rapidity:
\eqn\IIxxvii{
p_0 (\th ) = m \cosh (\th ) ~~~~~~~~p_1 (\th ) = m \sinh (\th ) , }
and $T \ket {\pm 1/2 , \th } = \pm \ket{\pm 1/2 , \th } $.
A set of chiral fields of topological charge $\pm 1$ with
non-vanishing matrix elements between the states and the vacuum
can be taken to be
\eqn\IIxxviib{\eqalign{
 \Psi_{\pm}(x,t)\ &=\ \exp\(\pm\frac{i}{\hb}\ \vphi(x,t)\)\cr
 \Psib_{\pm}(x,t)\ &=\ \exp\(\mp\frac{i}{\hb}\ \vphib(x,t)\) .\cr  }}
The fields $\Psi_\pm$ and $\Psib_\pm$ do not create independent
particle states, for the usual reasons\foot{e.g. a free Dirac
fermion in 2 dimensions has 4 components, which create 2 independent
particle states.}.
The representation of \slqh on the space of one-particle states
can be shown to be
\eqn\IIxxviii{\eqalign{
Q_{\pm}\ &=\ c\ e^{\th/\ga}\ \sigma_{\pm}\ q^{\pm \sigma_3 /2} \cr
\bQ_{\pm}\ &=\ c\ e^{-\th/\ga}\ \sigma_{\pm}\ q^{\mp \sigma_3 /2} \cr
T &= \sigma_3 ,  \cr}}
where $\sigma_\pm , \sigma_3$ are the Pauli spin matrices\foot{
$\sigma_3 = \left( \matrix{1&0\cr 0 &-1\cr} \right)$,
$\sigma_+ = \left( \matrix{0&2\cr 0 &0\cr} \right)$,
$\sigma_- = \left( \matrix{0&0\cr 2 &0\cr} \right)$}.
The on-shell operators $\exp (\th /\ga )$ are a consequence of the
Lorentz spin of the conserved charges, since the Lorentz boost
generator is represented as $-\d_\th $ on-shell.

The above representation of the \slqh loop algebra is in the so-called
principal gradation.  Gradations of loop algebras are only
meaningful mathematically up to inner automorphisms.  The
standard principal gradation of the \slh loop algebra
is $E_1^+ = x E_+ , E_1^- = x^{-1} E_-$, $E_0^+ = x E_- ,
E_0^- = x^{-1} E_+ $, $H_1 = -H_0 = H$, where $E_\pm , H$ generate
the finite dimensional $sl(2)$ Lie algebra, and $x$ is a
`spectral' parameter.  Another gradation we will make use of
is the homogeneous one, defined by the automorphism
\eqn\auto{
\sigma^{-1} a_{ {\rm prin.} } (x) \sigma = a_{ {\rm homo.} } (x^2 ),}
where
$\sigma = x^{H/2}$, and $a(x) \in \hat{sl(2)}$.  In this homogeneous
gradation one has $E_1^+ = E_+, E_1^- = E_-$, $E_0^+ = x^2 E_- ,
E_0^- = x^{-2} E_+$.  For our particular application,
$x$ is identified with $\exp (\th / \ga )$ and is a consequence of the
Lorentz spin of the conserved charges.  Thus, though the choice of
gradation is not intrinsically meaningful mathematically, different
gradations are certainly to be distinguished physically, since they
reflect scaling dimensions of the conserved currents involved.

\def\S{{\hat{S}}}
\def\x#1{{ e^{\th_#1 / \ga} }}
\def\xm#1{{ e^{-\th_#1 / \ga} }}

The action of the conserved charges on the multiparticle states is
provided by the comultiplication which follows from \IIxxiii\ and
the braiding of the non-local currents with the soliton fields.
This comultiplication derived in the quantum field theory is
equivalent to \IIix .  The two-particle to two-particle
S-matrix $\S$ is an operator from $V_1 \ot V_2$ to $V_2 \ot V_1$,
where $V_i$ are the vector spaces spanned by $\ket{ \pm 1/2 , \th_i }$.
The \slqh symmetry of the S-matrix is the condition
\eqn\IIxxixa{
\S_{12}
(\th_1 - \th_2 ;q ) \,\De_{12} (a) = \De_{21}
(a) \, \S_{12}  (\th_1 -\th_2 ;q) ~~~~~~a\in
\CU_q \( \hat{sl(2)} \) .}
Explicitly:
\eqn\IIxxix{\eqalign{
&\BBL \S (\th ;q)\ , \sigma_3 \otimes 1 + 1\otimes \sigma_3 \ \BBR =0 \cr
\S \(\th ;q\)&\
\( \x 1 \sigma_\pm q^{\pm \sigma_3 / 2} \ot 1
+ q^{\pm\sigma_3} \ot \x 2 \sigma_\pm q^{\pm\sigma_3 /2} \) \cr
&= \( \x 2 \sigma_\pm q^{\pm\sigma_3 /2} \ot 1 + q^{\pm \sigma_3} \ot
\x 1 \sigma_\pm q^{\pm\sigma_3 /2}  \) \ \S \( \th; q \) \cr
\S \(\th ; q \) &\
\( \xm 1 \sigma_\pm q^{\mp \sigma_3 / 2} \ot 1
+ q^{\mp\sigma_3} \ot \xm 2 \sigma_\pm q^{\mp\sigma_3 /2} \) \cr
&= \( \xm 2 \sigma_\pm q^{\mp\sigma_3 /2} \ot 1 + q^{\mp \sigma_3} \ot
\xm 1 \sigma_\pm q^{\mp\sigma_3 /2}  \) \ \S \( \th; q \) .
 \cr} }
 ($\th = \th_1 - \th_2$.)
The minimal solution to these symmetry equations is the conjectured
S-matrix of SG solitons\rZamoi .

Readers familiar with the QISM will recognize the above equations as
those which
characterize  the $R(x,q)$ matrix for the SG model, except for the
identification of the spectral parameter $x$ and the parameter $q$.
Indeed, solutions to the above equations were first obtained in
\rSTF .  However in the QISM $R(x,q)$ describes the commutation
relations of the monodromy matrix, thus $R(x,q)$ is not a priori
related to the S-matrix;  in fact in the QISM $x$ is a formal parameter, and
q is different from \eqd .  The non-local conserved charges provide
an explanation for why $\S$ is related to $R$, and fixes the non-perturbative
dependence on the coupling $\bh$.
\bigskip
\noindent
2c. {\it Deformed Serre Relations}
\medskip

\def\vl{\vartheta}
\def\dis{\displaystyle}

In this subsection we show how the deformed Serre relations are
derived from braiding relations and from the knowledge of the
ultraviolet behavior of the model.  Let us consider the general
setting of the preceding subsection, but with `abelian' braiding.
That is, we suppose to have a conserved non-local current $J_\mu (x,t)$
with braiding relations
\eqn\seri{
J_\mu (x,t )\,  J_\nu (y,t) = e^{i\pi \vl} \> J_\nu (y,t) \, J_\mu (x,t) ,
{}~~~~~x<y , }
and a field $\Phi$ such that
\eqn\serii{
J_\mu (x,t) \, \Phi (y,t) = e^{i\pi \eta} \> \Phi (y,t) \, J_\mu (x,t) ,
{}~~~~~x<y . }
Let $Q$ be the conserved charge corresponding to $J_\mu$.  The
expression
\eqn\seriii{
{\rm ad}_Q^{\dis ~n} \( \Phi (y,t) \) = \inv{(2\pi i)^n} \int \prod_{j=1}^n
dz^\nu_j \, \ep^{\nu\mu} \, J_\mu (z_j ) \ \Phi (y,t) }
is a multiple integral over a product of loops going from
$-\infty$ to $-\infty$ around $y$.  If this integration domain
could be shrunk to a region close to $y$, as one can do in the case
of local currents, then we would use our knowledge of the short
distance behavior of the theory to do the computation.
When is this possible?  The answer can be formulated in terms
of twisted homology \rFW , but here we choose a more direct
approach.  Let us replace the point $-\infty$ where the
integration contours originate and end by a point $(x,t)$ with
the same time coordinate as $(y,t)$ (figure 2),
and ask when the resulting integral \seriii\ is independent of
$x$.

\vskip 2cm
\ \ \ \ \ \ \ \ \ \ \ \ \ \ \ \ \ \ $\cdot\ (y,t)$ \ \ \ \ \
\ \ \ \ \ \ \ \ \ \ \ \ \ \ \ \ \ \ \ \ \ \ \
\ \ \ \ \ \ \ \ \ \ \ \ \ \ $\cdot\ (x,t)$
\vskip 2cm
\centerline{\eightrm Fig. 2. The contour of integration
in \seriii\ with endpoint $(x,t)$.}
\vskip 10pt

\noindent
Let us compute the derivative with respect to $x$.
Both upper and lower integration bounds of the integral over
$z_j$ are equal to $x$.  Thus the derivative with respect to
$x$ is a sum of boundary terms, two for each integration variable,
which are all equal up to a phase to
\eqn\seriv{
J_t (x,t) \int \prod_{j=1}^{n-1} \, \(J^\mu (z_j ) \ep_{\mu\nu}
dz_j^\nu \) \, \Phi (y,t) . }
The phase of the lower integration bound of the j-th variable
is obtained by braiding $J^\mu (z_j )$ through
$J(z_1 ) \cdots J(z_{j-1} ) $ and is thus $e^{i\pi (j-1) \vl } $.
Similarly, we get the phase $e^{2\pi i (n-j) \vl + 2\pi i \eta }
e^{i\pi (j-1) \vl}$ from the upper integration bound. Summing up, the
derivative with respect to $x$ is equal to \seriv\ times
\eqn\serv{
\( e^{2\pi i \eta + i\pi \vl (n-1)} -1 \)
\frac{e^{i\pi \eta\vl} - 1 }{e^{i\pi \vl} -1 } . }
We conclude that if the braiding phases are related by the
formula
\eqn\servi{
\eta + \vl \> \frac{n-1}{2} = 0 ~~{\rm mod} ~1, }
the integration domain can be shrunk to a product of loops as in
figure 2 with $x$ arbitrarily close to $y$.

We can thus argue using the short distance behavior of the theory.
Suppose that $J,\Phi$ have ultraviolet scaling dimension
$h_J , h_\Phi$.  The field
${\rm ad}_Q^{\displaystyle ~n} \( \Phi \) $
has
then scaling dimension
\eqn\servii{
n(h_J - 1 ) + h_\Phi . }
This information, plus the knowledge of the quantum numbers
of the operators, can be sufficient to identify
${\rm ad}_Q^{\dis ~n} \( \Phi \) $ with a field in the conformal
field theory describing the ultraviolet behavior.  In particular
if no operator exists with the given quantum numbers and scaling
dimension \servii , we conclude that
${\rm ad}_Q^{\dis ~n} \( \Phi \) = 0$.

\def\a{{\rm ad}}
Let us apply these ideas to the Serre relations.
They are in this case
\eqn\serviii{
{\rm ad}_{Q_\pm}^{\dis ~3} \( Q_\mp \) = {\rm ad}_{\Qb_\pm}^{\dis ~3}
\( \Qb_\mp \)
= 0 . }
Consider the expression
$\a_{Q_+}^{\dis ~n} \( J^-_\mu (y,t) \) $. We see
that we are in the situation described above with
\eqn\serix{
\vl = -\frac{4}{\bh^2} ~~~~, ~~~~\eta= \frac{4}{\bh^2}
}
and condition \servi\ is precisely satisfied when $n=3$.
Shrinking the integration domain we can reduce the computation
to free conformal field theory: the scaling dimension of the
component of
$\a_{Q_+}^{\dis ~3} \( J^-_\mu (y,t) \)$ of leading order in $\lambda$ has
ultraviolet scaling dimension $8/\bh^2 - 3$. On the other hand
the result is in the conformal family of the vertex operator
$\exp (i4 \vphi / \bh )$.  But the fields in this family
have scaling dimension equal to $8/\bh^2$ or higher.  We conclude
that to leading order in $\lambda$,
$\a_{Q_+}^{\dis ~3} \( J^-_\mu (y,t) \)  = 0$.
The terms of higher order in $\lambda$ are treated in a similar
way:  the conformal weights are in all cases lower that the
minimal values in the corresponding conformal families, and we
have
\eqn\serxi{
\a_{Q_+}^{\dis ~3} \( J^-_\mu (y,t) \) = 0, }
implying the first Serre relation \serviii .  The other relations in
\serviii\ are  proven in the same way.

\def\vPhi{{\vec{\Phi}}}
The above proof of the Serre relations extends easily to an arbitrary
quantum affine Lie algebra.  Let $\vec{\Phi}$ denote a vector of fields
valued in the Cartan subalgebra of an arbitrary affine Lie algebra
$\Gh$, and consider the affine Toda theory:
\eqn\Vi{
S= \inv{4\pi} \int d^2z\ \d_z \vPhi \cdot \d_\zb \vPhi
+ \frac{\la }{2\pi }\int d^2 z \sum_{\av_j\ {\rm simple}}
\exp \( {-i\be \frac{2}{|\av_j|^2} \av_j\cdot\vPhi } \) .}
This theory has non-local conserved charges $Q_j$
generated by the purely chiral components\nlc
\eqn\serxiii{
J_{\av_j} (x) = \exp \( \frac{i}{\be} \av_j \cdot \vec{\vphi} \) , }
and also charges $\Qb_j$ generated by the purely anti-chiral
components
\eqn\serxiii{
\Jb_{-\av_j} (x) = \exp \( \frac{i}{\be} \av_j \cdot \vec{\vphib} \) . }
($\vec{\vphi}$ and $\vec{\vphib}$ are quasi-chiral components of $\vPhi$
as in \IIxiv .
Define the topological charges
\eqn\top{
H_i = \frac{\be}{2\pi i} \int dx \,  \av_i \cdot  \d_x \vPhi (x,t) . }
Using the braiding relations
\eqn\serxiv{
J_{\av_j} (x) \, \Jb_{-\av_j} (y)
=
q^{-|\av_j |^2} \> \Jb_{-\av_j} (y) \, J_{\av_j} (x) ~~~~~\forall\ x,y ,}
where $q=\exp (-i\pi / \be^2 )$, and results from \nlc ,  one can show
that these charges satisfy  the \UGm algebra in the form \IIx .

Consider now the deformed Serre relations \IIxi .  From
\eqn\serxv{
J_{\av_i} (x) \, J_{\av_j} (y)
=
\exp \( {-\frac{i\pi}{\be^2} \av_i \cdot \av_j } \)
\> J_{\av_j} (y) \, J_{\av_i} (x) ~~~~~x<y ,}
one sees that we can apply the general result described above with
\eqn\serxvi{
\vl = - \frac{|\av_i |^2 }{\be^2} ~~~~, ~~~~\eta = - \frac{\av_i \cdot \av_j}
{\be^2} . }
The solution to \servi\ is thus $n=1-a_{ij}$, where $a_{ij}$ is the
generalized Cartan matrix of $\Gh$.  Using scaling arguments similar to the
\slqh case, one establishes \IIxi .

\def\Ps#1#2#3{{ \Psi^{(#1)}_{#2#3} }}
\def\cft{ {\rm CFT} }
\def\CHb{ {\bar{\CH}} }
\def\pert{ {\rm pert.} }
\def\p#1{ {\phi^{(#1)}} }
\def\pb#1{ {\phib^{(#1)}} }
\def\P#1{ {\Phi^{(#1)} } }

\newsec{Superselection Sectors in Massive Perturbations of CFT}

As we will see, the proper general framework for understanding
certain features of the RSG quantum field theory is the
general theory of superselection sectors, in particular the
case of non-abelian sectors, as outlined in \rFroh\rFRS .
The basic ingredients of a theory with superselection
sectors are as follows.  The complete Hilbert space $\CH$ of the
theory is decomposed into  `charge' sectors $\CH_i : \CH =
\sum_i \CH_i $.  Fields $\Psi (x)$ are operators from
$\CH \to \CH$.  In particular, the charged fields
$\Psi^{(k)}_{ji} (x)$ intertwine the spaces $\CH_i$ and
$\CH_j$:
\eqn\sa{
\Ps kji (x): ~~~~\CH_i \to \CH_j . }

The superselection sectors of a CFT can be summarized as follows.
One has chiral and anti-chiral primary fields $\phi^{(i)} (z)$ and
$\phib^{(i)} (\zb )$\foot{We follow the convention that
$\p i , \pb i $ refer to chiral and anti-chiral fields respectively,
whereas the local spinless fields are $\Phi^{(i)} = \p i \pb i $. },
 and their associated states
$\ket{i} = \p i (0)\ket{0}, ~ \ket{\bar{i}} = \pb i (0) \ket{0}$.
The sectors $\CH_i$ ($\CHb_i$) correspond to the states
$\ket{i}$ ($\ket{\bar{i}}$) and their descendents with respect to
the chiral algebra.  The complete Hilbert space is given by
\eqn\sb{
\CH^\cft = \sum_{i,j} \CH_i^\cft \ot \CHb_j^\cft . }
Note that we do not assume that CFT consists only of local
fields, which corresponds to
$\CH^\cft = \sum_i \CH_i^\cft \ot \CHb_i^\cft$.
This is important for massive perturbations of CFT, since non-local
fields are generally required for example to create particle states in the
charge sectors.  Let $N^{ij}_k $ denote the fusion algebra of the
CFT: $ i \times j = \sum_k N^{ij}_k  k $, and similarly for
the anti-chiral fields.  For every non-zero $N^{ij}_k$ there exists
an intertwiner (or chiral vertex operator in this context, see
\ref\rTK{\TK}\ref\rMS{\MSi}\ref\rFFK{\FFK}\ for precise definitions):
\eqn\sc{
\phi^{(i)}_{kj}: ~~\CH_j^\cft \to \CH_k^\cft ~~~~~; ~~~~~
\phib^{(i)}_{kj}: ~~ \CHb_j^\cft \to \CHb_k^\cft . }

Consider now a massive perturbation of the CFT, formally
described by the action
\eqn\sd{
S\ =\ S_{{\rm CFT}} + \frac{\la}{2\pi }
	\int d^2z\ \Phi_\pert  (z,\zb) .}
The perturbing field $\Phi_\pert (z,\zb )$ is a local spinless
field, and can be generally expressed as $\Phi_\pert
= \phi_\pert \phib_\pert$, where $\phi_\pert$ ($\phib_\pert$)
is a (anti-)chiral field in the CFT,
with scaling dimensions $h(\phi_\pert )= \bar{h} (\phib_\pert )$.
The superselection sectors of the perturbed CFT are generally
fewer than those of the CFT. We propose their following
characterization:

{\it The superselection sectors of the perturbed CFT are intertwined
by the set of  fields which are local with respect to the
perturbing field $\Phi_\pert$.  }

One is led to the above characterization by the following
considerations. Let $\CF$ denote the space of fields of the
CFT, which includes all possible, not necessarily spinless,
products of chiral with anti-chiral fields,
 and $\CF_L \subset \CF$ the space of fields which are
local with respect to $\Phi_\pert$.  From general principles, one
expects that the intertwining fields should be local
with respect to observables.  For the perturbed CFT \sd , the
trace of the energy-momentum tensor is equal to
$\be (\lambda ) \Phi_\pert$, where $\be (\lambda )$ is the beta-function
\ref\rZrg{\Zflow}.  Thus locality with respect to the energy-momentum
tensor requires the intertwining fields to be in $\CF_L$.
Alternatively, one may argue as follows. Let $\Psi (x) \in \CF$ and
consider a correlation function involving this field in perturbation
theory:
\eqn\se{
\langle \Psi (x) \cdots \rangle = \langle \Psi (x) \cdots \rangle_\cft
+ \lambda \int d^2 z \vev{\Phi_\pert (z,\zb ) \Psi (x) \cdots }
+ O(\lambda^2 ) + \ldots }
Unless $\Psi (x) \in \CF_L$, the integral over $z,\zb$ is not well-defined.
Indeed, if $\Psi (z)$ is a chiral field in $\CF_L$, then
$\d_\zb \Psi$ is completely well-defined and computable in
perturbation theory\rzamoiii .

In the language of algebraic quantum field theory, the concept
of superselection sectors is based on the choice of an
observable algebra of (bounded functions of) local fields,
of which the superselection sectors are representation
spaces. Intertwining fields are then local with respect
to the observable algebra.
In conformal field theory, it is natural to choose as
observable algebra the algebra
generated by the energy momentum tensor, or possibly some
larger local chiral algebra. When one considers perturbations
of conformal models by some local operator, parametrized by
a coupling constant $\lambda$, the above reasoning shows that
one should adjoin the perturbing field to the observable
algebra. One effect of this procedure is, as we have seen, that
some sectors of the CFT disappear, because their intertwining fields
are not local with respect to the perturbation. Another effect
is that some CFT sectors are welded because the larger algebra
does not leave them invariant.

In a theory in which we have a description in
terms of microscopic degrees of freedom (e.g. the field $\Phi$
in the SG model or the order parameter in a Landau-Ginzburg
theory) one has an absolute notion of locality: local
fields at a point $x$ are functions of the microscopic degrees
of freedom in an infinitesimal neighborhood of $x$.
In order to describe the topological behaviour of the model
one should not distinguish between sectors which are intertwined
by local fields. Let us call the larger sectors obtained this
way topological sectors. A good example where the distinction can
be made is the free massless bosonic field. Sectors are labeled
by two numbers, the magnetic (topological) charge and the
electric charge. In each sector we have a different representation
of the current algebra generated by $\epsilon_{\mu\nu}
\partial^\nu\Phi$ and  $\partial_\mu\Phi$. The intertwining
fields for the electric charge are exponentials of the local
field $\Phi$ whereas the intertwining fields carrying magnetic
charge are non local. Thus we see that the topological sectors
are labeled by the magnetic charge only, so that the sectors
of different electric charge but with the same magnetic charge are
welded into a single topological sector.

Let us specify the sectors of the perturbed theory more precisely.
Define an index set $L$, such that for $i\in L$,
$\p i (z)$ is local with respect to $\phi_\pert (z)$ in the
CFT. Denote by $\p {\bar i}$ the field conjugate to
$\p i$.
  The field $\pb i (\zb ), i\in L$ is also local with respect
to $\phib_\pert (\zb )$.   Thus $\CF_L$ is spanned by
$\p i , \pb i , i\in L$, and any product of these fields with
 local fields.  We consider the diagonal theory in which the local
primary fields are $\p i\pb i$.
The sectors corresponding to
fields $\p i$ and $\pb {\bar i}$ are intertwined by
local fields, thus they do
not intertwine distinct topological
sectors;  the chiral and anti-chiral CFT sectors
are thus welded.
The topological sectors of the perturbed theory
are thus
\eqn\sf{
\CH^\pert = \sum_{i\in L} \CH_i^\pert , }
and they are intertwined as follows
\eqn\sg{
\phi^{(i)}_{kj} , ~\phib^{(i)}_{kj} : ~~~\CH_j^\pert \to \CH_k^\pert , ~~~~~~
i,j,k \in L . }
Since braiding is a property of the short distance behavior of a
quantum field theory, the braiding of the intertwiners in the CFT and
in the perturbed theory are the same.

Let us illustrate these ideas by using them to re-derive the
superselection sectors of the SG theory.  Fr\"ohlich established
the existence of sectors of integer topological charge \IIxiii\
from the general principles of algebraic quantum field theory\rFroh ;
one important point of this work is that it did not rely on any
semi-classical arguments, as opposed to \rmand .
Since our characterization of sectors is rather
different, it is important to check it in this simple example.
The chiral primary fields of the $c=1$ free boson theory are
$e^{i\al \vphi (z)}$, which defines the sector $\CH_\al^\cft$, where
$\al$ is any real parameter.  The perturbing field is $\cos (\bh \Phi )$.
{}From the operator product expansion
\eqn\sh{
\exp \( {i\frac{n}{\bh} \vphi (z) } \)  \>
\exp \( {-i \bh \vphi (w) } \)
= \inv{ (z-w)^n }
\exp \( { i (\frac{n}{\bh} - \bh ) \vphi (w)} \)  + \ldots }
one deduces that $e^{in\vphi /\bh} \in \CF_L$, for $n$ an integer.
Thus the sectors of the SG theory are $\CH^{SG} = \sum_{n\in \Z}
\CH_n^{SG}$.   From \IIxivc\ one sees that the fields
$e^{in\vphi /\bh}$ and $e^{-in\vphib /\bh}$  have precisely topological
charge $n$.  They intertwine the sectors according to $\CH_m \to \CH_{m+n}$.
Thus our conclusions are in complete agreement with the results in
\rFroh . Indeed, the above analysis provides an alternative
derivation of the normalization of the topological current \IIxiiib .

We turn now to the sectors of the RSG theory.  This theory is
a perturbation of the minimal unitary series by the operator $\Ph 13$.
The chiral primary fields of the minimal model are $\ph nm (z)$ with
$1\leq n \leq p-1, 1\leq m \leq p$, with conformal dimension
\eqn\si{
h_{nm} = \frac{ [(p+1)n - pm ]^2 -1 }{4p(p+1)} . }
{}From these dimensions one finds that $h_{n,1} + h_{1,3} - h_{n,3} = n-1$,
thus the fields $\ph n1$ are in $\CF_L$.  relabel the integers
$n$ of $\ph n1$ as $n=2j+1$ and let $\ph n1 \equiv \p j$.  The
superselection sectors of the RSG theory are thus
\eqn\sk{
\CH^{RSG} = \sum_{j \in \{ 0,1/2, 1, ..., p/2 -1 \} } \CH_j^{RSG} .}
The fusion rules are the standard ones
\eqn\efus{
j_1 \times j_2 = \sum_{j=|j_1 -j_2 |}^{ {\rm min} (j_1 + j_2 ;
p-2-j_1 - j_2 )}  j  }
Let us relabel the intertwiners $\phi^{(n,1)}_{(n_2 ,1)(n_1 ,1)} $ in
a similar way as $\phi^{(j)}_{j_2 j_1 } $.  Then the sectors are
intertwined as follows
\eqn\sint{
\phi^{(j)}_{j_2 j_1} , ~~\phib^{(j)}_{j_2 j_1} : ~~~~~
\CH_{j_1} \to \CH_{j_2} . }

In the SG theory, the soliton states are created by the intertwining
fields.  Thus on general grounds, one expects qualitatively that the spectrum
of massive kink states in the RSG theory should intertwine the
sectors $\CH_j$.  This will be made explicit in the next section.

As a further example, consider the $su(2)$ Wess-Zumino-Witten model
at level $k$ perturbed by the marginal operator $\Phi_{\rm pert.}
= \sum_a J^a \bar{J}^a$, where $J^a, \bar{J}^a$ are the chiral
and anti-chiral $su(2)$ affine currents. This is a massive model due to
its non-zero beta-function (for $k=1$ this is the chiral Gross-Neveu
model).
Since all primary fields are local with respect to $J^a$, the sectors
of the perturbed theory in this case are just a welding of chiral
and anti-chiral CFT sectors:
$\CH^{\rm pert.} = \sum_{j\leq k/2} \CH^{\rm pert.}_j$.
The conjectured spectrum of these models\rfssg\rABL\  indeed reflects the
existence of these sectors.

\bigskip

\newsec{The Restricted Quantum Affine Symmetry}
\medskip

In this section we will describe the restriction of the SG
theory from the vantage of the quantum affine symmetry described
above.
We will show that the restriction
of the SG theory can be described as the restriction to fields
and
asymptotic states  that form a quotient of the space
of singular vectors with respect to an
\slq subalgebra of \slqh .
We will describe the restriction in two steps.  First we construct
fields and states which are covariant with respect to \slq .
Then we project onto  the restricted space of
fields and states. We then describe how the residual \slqh symmetry
acts on the restricted model.

\medskip\noindent
4a. {\it Quantum Group Covariant Fields and States}

\medskip

The \slqh algebra has two \slq subalgebras generated by
$\{ Q_+, \Qb_- , T \}$ and $\{ Q_- , \Qb_+ , T \}$.  (These two
subalgebras are not independent.)
The energy-momentum tensor of the SG theory is not invariant under
\slq , as first noticed by
Reshetikhin and Smirnov \rRS . To see this explicitly,
for simplicity consider the conformal limit,
where $T_{zz} = - \d_z \vphi \d_z \vphi /2$ and similarly for
$T_{\zb\zb}$.  Then from explicit computation one sees that
$[Q_+ , T_{zz} (z) ] \neq 0$.
However familiar results from the Feigin-Fuchs construction \ref\rDF{\Dotsenko}
\ref\rfeld{\Felderi} , indicate that $T_{zz}$ can be made invariant with
respect to $Q_+$ by modifying it to include a background charge term:
\eqn\IIIi{ T_{zz}'  \ =\ -\half\d_z\vphi\d_z\vphi +
	i\sqrt{2}\alpha_0\ \d_z^2\vphi .}
By choosing $\al_0$ appropriately such that $Q_+$ becomes identified with
a screening operator, one can ensure that
\eqn\IIIii{
\[ Q_+ , T_{zz} ' \] = 0. }
The condition that $Q_+$ is a screening operator is equivalent to the
requirement that it have scaling dimension 0.  The operators
$e^{i\al \vphi } $  now have
dimension $\al^2 /2 - \sqrt{2} \al \al_0 $ with respect to $T_{zz}'$.
Thus $\al_0$ is fixed to be a solution of
\eqn\IIIiii{ 1\ =\ \frac{2}{\hb^2}-\frac{2\sqrt{2}}{\hb}\ \al_0~~.}

As usual the background charge term in $T_{zz} '$ contributes to the
conformal anomaly $c= 1-24\al_0^2$.
For
\eqn\IIIiv{
\frac{\bh}{\sqrt{2}} = \sqrt{ \frac{p}{p+1} } , }
$\al_0$ is such that $c$ is that of the minimal unitary series \Ii .
A similar analysis applies to $\Qb_-$ and $T_{\zb\zb}$.  The above
inclusion of a background charge does not affect the fact that
$Q_+ , \Qb_- , T$ generate the algebra \slq . Note that it is not
possible to simultaneously demand $T_{zz} '$ to be invariant under
both \slq subalgebras, since $Q_+$ and $Q_-$ cannot simultaneously
be given dimension 0.  For the remainder of this section, \slq will
refer to the subalgebra generated by $Q_+, \Qb_- $, and $T$.
We also rescale the charges $Q_+ , \Qb_- \to c^{-1} Q_- , c^{-1} \Qb_-$,
where the constant $c$ is defined in \IIxxvi , so that the
resulting charges satisfy the usual \slq relations.

In order to make certain arguments, let us follow the Feigin-Fuchs
construction and define the vertex operator fields
\eqn\IIIv{
V_{nm} (x) = e^{i\sqrt{2} \al_{nm} \vphi } ~~, ~~~~~
\bar{V}_{nm}  (x) = e^{i\sqrt{2} \al_{nm} \vphib } ~~~~~~, }
where
\eqn\IIIvi{\eqalign{
\al_{nm} &= \half (1-n) \al_+  + \half (1-m) \al_-  \cr
\al_+ &= \frac{\sqrt{2}}{\bh} , ~~~~~~\al_- = - \frac{\bh}{\sqrt{2}}
\cr
1&\leq n \leq p-1 , ~~~~~~~1\leq m \leq p .\cr} }
These fields represent the minimal model primary fields
$\ph nm , \phb nm$.
The currents $J_-$ and $\Jb_+$ in \IIxvi{}\ are associated with
the primary fields $\ph 31$ and $\phb 31$.  Note also that the
soliton fields $\Psi_-$ and $\bar{\Psi}_+$ are associated with
$\ph 21$ and $\phb 21$, whereas $\Psi_+$ and $\bar{\Psi}_-$ are
not in the usual spectrum of primary fields.
Note that of the two operators $\exp (\pm i \bh \Phi )$
that define $\cos (\bh \Phi )$ in \IIxii ,
one becomes a screening operator, the other becomes the field
$\Phi^{(1,3)}$ of dimension $(p-1)/(p+1)$, so that we are actually
describing the model \Iib .

\def\p#1#2{\phi^{(#1)}_{#2}}
\def\pb#1#2{\phib^{(#1)}_{#2}}

Define the fields $\p j{-j} (x)$ and $\pb jj (x)$ as
\eqn\IIIvii{\eqalign{
\p j{-j} (x) &\equiv V_{2j+1,1} (x) = \exp \(
{- \frac{2i}{\bh} \> j \> \vphi (x) } \)  \cr
\pb j{j} (x) &\equiv \bar{V}_{2j+1,1} (x)
= \exp \( {- \frac{2i}{\bh} \>  j \> \vphib (x) } \)  . \cr}}
Now consider the fields obtained by adjoint action with
$Q_+$ and $\Qb_-$:
\eqn\IIIviii{\eqalign{
\p j{-m} (x) & = \a_{Q_+}^{j-m} \( \p j{-j} (x) \) \cr
\pb j{m} (x) & = \a_{\Qb_-}^{j-m} \( \pb j{j} (x) \) . \cr}}
The adjoint action in \IIIviii\ is defined in the quantum field
theory as an integral of the current $J_+$ or $\Jb_-$ along a
contour surrounding $x$, as in \IIxx .  These fields were
considered by Gomez and Sierra in their study of the quantum
group symmetry of minimal conformal models and shown to
comprise spin-j \slq multiplets.  Related results can be
found in \rFW .  We will establish these
properties using other arguments.

It is not difficult to see from the results of section 2c that
through the adjoint action \IIIviii\ one indeed obtains  a finite
number of fields.  Namely, from the braiding relations
\eqn\IIIix{\eqalign{
J_+ (x) \,  J_+ (y) &= e^{-4\pi i /\bh^2 } \> J_+ (y) \, J_+ (x) ~~~~~x<y
\cr
J_+ (x) \, \p j{-j} (y) &= e^{j\, 4\pi i / \bh^2 } \>
\p j{-j} (y) \, J_+ (x) , \cr }}
and the general condition \servi , one finds that contours defining
$\p j{j+1}$ can be closed.  Then using arguments as for establishing
the Serre relations one finds
\eqn\IIIx{
\p j{j+1} (x) = \pb j{-j-1} (x) = 0 . }
Thus the fields $\p jm , \pb jm , -j\leq m \leq j$ define
$2j+1$ dimensional multiplets.

We now consider the transformation properties of the fields
$\p jm, \pb jm$ with respect to \slq .
{}From the braiding relations
\eqn\IIIxi{
J_+ (x) \, \p j{-m} (y) = q^{-2m} \p j{-m} (y) \, J_+ (x) ~~~~~x<y ,}
and \IIxxi\ one finds
\eqn\IIIxii{
\a_{Q_+} \( \p j{-m} (x) \) = Q_+ \p j{-m} (x)
-q^{-2m} \p j{-m} (x) Q_+ . }
Using the fact that the field $\p j{-m}$  has topological charge
$T= -2m$, \IIIxii\ can be expressed as
\eqn\IIIxiii{
\a_{Q_+} \( \p j{-m} (x) \) = Q_+ \p j{-m} (x)
+ q^T  \p j{-m} (x) s\( Q_+ \)  , }
where the antipode of $Q_+$ is given in \IIix :
$s(Q_+ ) = -q^{-T} Q_+$.

The adjoint action expressed in the form \IIIxiii\ has some
important properties which we now explain\rLS\rBF .
Consider an arbitrary Hopf algebra $\CA$ with the properties
\IIiii{} .  Let $a\in \CA$ and express its comultiplication as in
\IIv . Define the adjoint action on a field or product of
fields as
\eqn\IIIxiv{
\a_a \bigr( \Phi (x_1 ) \cdots \Phi (x_n ) \bigl) = \sum_i
a_i \> \Phi (x_1 ) \cdots \Phi (x_n ) \> s(a^i ) . }
Then

\noindent
1. Fields related through adjoint
action form a representation of $\CA$.  More precisely let
$\Phi_v (x)$ denote the set of fields so obtained, where $v$ spans
a vector space $V$, and let $\rho_V (a)$ denote a representation of
$\CA$ on $V$.  Then
\eqn\IIIxv{
\a_a \( \Phi_v (x) \) = \Phi_{\rho_V (a) v} (x). }

\noindent 2.
\eqn\IIIxvi{
\a_a \bigl( \Phi_{v_1} (x_1 ) \> \Phi_{v_2} (x_2 ) \bigr)
= \sum_i \a_{a_i} \bigl( \Phi_{v_1} (x_1 ) \bigr)
\a_{a^i} \bigl( \Phi_{v_2} (x_2 ) \bigr) . }

\def\Rb{\bar{R}}

\noindent
These properties are a consequence of the Hopf algebra properties \IIiii{}.
  The relations \IIIxv\IIIxvi\  apply as well to spacetime independent
operators such as conserved charges.

Thus the fields $\p jm$ indeed form $2j+1$ dimensional representations
of {\slq}.  Similar results apply to the fields $\pb jm$.  We will need
the braiding relations of these fields.  It follows from general
principles that
\eqn\IIIxvii{\eqalign{
\p {j_1}{m_1} (x_1 ) \ \p {j_2}{m_2} (x_2) &= \sum_{m'_1 , m'_2}
\( R^{j_1 j_2} \)^{m_1 'm_2 '}_{m_1 m_2} \
\p {j_2}{m_2 '} (x_2 ) \ \p {j_1}{m_1 '} (x_1)  ~~~~~x_1 <x_2 \cr
\pb {j_1}{m_1} (x_1 ) \ \pb {j_2}{m_2} (x_2) &= \sum_{m'_1 , m'_2}
\( \Rb^{j_1 j_2} \)^{m_1 'm_2 '}_{m_1 m_2} \
\pb {j_2}{m_2 '} (x_2 ) \ \pb {j_1}{m_1 '} (x_1)  ~~~~~x_1 <x_2 ,\cr }}
where $R^{j_1 j_2}$ is the universal $R$-matrix
\ref\rkol{\Kirillovi}\  for \slq evaluated in the
representations indicated\foot{The $R$-matrices, and also the q-Clebsch-Gordan
coefficients and q-6j symbols below, can be computed from the results in
\rkol .  Most of the relevant ones were evaluated explicitly in \rBLii .
}, and $\Rb = R^{-1}$.
The universal $R$-matrix has the defining relation
\eqn\IIIxviii{
R\, \De (a) = \De' (a) \, R }
where $\De ' = P \De$, and $P$ is the permutation operator.
The relations \IIIxvii\ are easily established by applying
$\a_a$ to both sides and using \IIIxvi\ to show that $R^{j_1 j_2}$
must satisfy its defining relations.   Alternatively these
braiding relations can be verified by explicit computation as
was done in \rGoS .  From \IIxivb{} one also has
\eqn\braid{
\pb {j_1}{m_1} (x_1 ) \ \p {j_2}{m_2} (x_2) = \sum_{m'_1 , m'_2}
\( R^{j_1 j_2} \)^{m_1 'm_2 '}_{m_1 m_2} \
\p {j_2}{m_2 '} (x_2 ) \ \pb {j_1}{m_1 '} (x_1)  ~~~~~\forall\ x_1 ,x_2
.}

\def\CJb{\bar{\CJ} }

We now apply these results to the RSG theory by first identifying
which of the above fields are especially meaningful in our context.
Since the charges $Q_\pm$ and $\Qb_\pm$ are conserved in the SG
theory, and we have the identifications $J_- = \p 1{-1} ,
\Jb_+ = \pb 11$, the fields
\eqn\ej{
\CJ_m (x) \equiv \p 1m (x) ~,~~~~~\CJb_m (x) = \pb 1m (x) }
generate some non-local conserved currents and charges associated
with the spin-1 representation of \slq . Namely,
\eqn\IIIxix{
\d_\zb \CJ_m = \d_z \CH_m ~~~, ~~~\d_z \CJb_m = \d_\zb \CHb_m
{}~~~~~m= \pm 1,0 }
for some $\CH_m , \CHb_m$,
and
\eqn\IIIxx{\eqalign{
Q^{(1)}_m &= \inv{2\pi i} \(  \int dz \CJ_m + \int d\zb \CH_m  \) \cr
\Qb^{(1)}_m &= \inv{2\pi i} \( \int d\zb \CJb_m + \int dz \CHb_m  \) \cr
}}
are conserved.
The Lorentz spin of these charges follows from the dimension of
the currents $J_- , \Jb_+$ (with respect to $T'$) and the fact that
$Q_+ , \Qb_-$ have spin 0:
\eqn\IIIxxb{
{\rm spin} \( Q^{(1)}_m \) = -{\rm spin} \( \Qb^{(1)}_m \) = 2/\ga = 2/p,}
where $\ga$ is defined in \IIxviii .  Note that the spin is
precisely doubled due to the inclusion of the background charge.

\def\Kb{\bar{K}}

The SG soliton fields $\Psi_\pm (x) , \Psib_\pm (x)$ do not
form spin-1/2 \slq multiplets.  Therefore we define new soliton
fields:
\eqn\IIIxxi{
K_m (x) \equiv \p {1/2}{m} (x) , ~~~~~\Kb_m (x) \equiv \pb {1/2}m (x) . }
These fields have topological charge $\pm 1$, however whereas
$K_{-1/2} = \Psi_-$ and $\Kb_{1/2} = \Psib_+$,
$K_{1/2} \neq \Psi_+ , \Kb_{-1/2} \neq \Psib_-$.  Recall that the
fields $\Psi_+ , \Psib_-$ were not in the spectrum of primary fields of
the minimal models; we have thus replaced them by more appropriate ones.

We define asymptotic states $\ket{K_m (\th )}$ to be created
by the fields $K_m (x)$ or $\Kb_m (x)$.  They can be normalized
such that
\eqn\IIIxxii{
\bra{0} K_m (x) \ket{K_{-m} (\th )} = \bra{0} \Kb_m (x)
\ket{ K_{-m} (\th ) } = e^{-ip(\th )\cdot x } . }

The S-matrix for the scattering of the states $\ket{K_m (\th )}$ can be
determined exactly by using its various non-local symmetries.  We
first determine the action of the charges $Q^{(1)}_m , \Qb^{(1)}_m$ on the
states.  The essential features of this action are a simple
consequence of the \slq symmetry.  Consider more generally a
Hopf algebra $\CA$.
Let $V'$ and $V''$ denote representations of $\CA$,
and $\CO$ an operator from $V'$ to $V''$: $\CO \in Hom(V' , V'').$
As usual $\CA$ acts on $\CO$ as
\eqn\IIIxxiii{
\a_a \( \CO \) = \sum_i a_i \, \CO \, s(a^i )  }
for $a\in \CA$ and $\De (a)$ of the form \IIv .  For all $v' \in V'$,
one has
\eqn\IIIxxiv{
a\, \CO \, v' = \sum_i \a_{a_i} \( \CO \) a^i \, v' . }
The above relation is easily proven using the properties \IIiii{} .
Namely, let $\De (a_i ) = \sum_j a_{ij} \ot {a_i}^j $
and $\De (a^i ) = \sum_j {a^i}_j \ot a^{ij} $.  The coassociativity
implies $\sum_{i,j} a_{ij} \ot {a_i}^j \ot a^i
= \sum_{i,j} a_i \ot {a^i}_j \ot a^{ij}$.  Thus the RHS of \IIIxxiv\
is
\eqn\IIIxxv{\eqalign{
\sum_i \a_{a_i} \( \CO \) \, a^i \, v' &=
\sum_{i,j} a_{ij} \, \CO \, s({a_i}^j ) \, a^i \, v' \cr
&= \sum_{i,j} a_i \, \CO \, s({a^i}_j ) \, a^{ij} \, v' . \cr}}
Now using the fact that $\sum_j s({a^i}_j ) a^{ij} = \ep (a^i )$ and
$\sum_i a_i \, \ep (a^i ) = a$, one has established \IIIxxiv .

Suppose that $\CO$ transforms in some representation $V$ of
$\CA$.  Let $V$ be spanned by vectors $v$, and denote the
operators $\{ \CO \}$ as $\CO_v$.  The operators $\CO_v$ are
maps $V\to Hom(V' , V'')$, and from the property analogous to
\IIIxv\ one has
\eqn\IIIxxvi{
\a_a \( \CO_v \) = \CO_{av} . }
The map $C$:
\eqn\IIIxxvii{
C:~~~V\ot V' \to V'' : v\ot v' \to \CO_v v' }
by \IIIxxiv\ is a homomorphism of $\CA$-modules.
More specifically
\eqn\IIIxxviii{\eqalign{
C\, \De (a) \, v\ot v' &= C\, a_i v \ot a^i v' \cr
&= \CO_{a_i v} \, a^i v' \cr
&= a \, C \, v\ot v' . \cr} }
Thus the map $C$ is, by definition,  a Clebsch-Gordan projector. In
the $sl(2)$ case it
is unique up to proportionality.  The results just described
may be thought of as a generalization of the Wigner-Eckart theorem
to an arbitrary Hopf algebra.

\def\j#1{ {j_#1} }
\def\cg#1#2#3#4#5#6{{
\left[ \matrix{{#1}&{#2}&{#3}\cr{#4}&{#5}&{#6}\cr} \right]_q
}}
\def\sj#1#2#3#4#5#6{{
\left\{ \matrix{{#1}&{#2}&{#3}\cr{#4}&{#5}&{#6} \cr} \right\}_q
}}

The above reasoning implies the following structure for matrix
elements:
\eqn\IIIxxix{\eqalign{
\bra{K_\be (\th )} \, Q^{(1)}_m \, \ket{K_\al (\th )}
&= \hat{c} \, e^{2\th /\ga} \cg 1{1/2}{1/2}m\be\al \cr
\bra{K_\be (\th )} \, \Qb^{(1)}_m \, \ket{K_\al (\th )}
&= \hat{c} \, e^{-2\th /\ga} \cg 1{1/2}{1/2}m\be\al \cr
}}
where $\hat{c}$ is some constant, and the brackets refer to
the \slq Clebsch-Gordan coefficients.
The factors $e^{\pm 2\th / \ga}$ are a consequence of \IIIxxb .

Consider now the action of the charges $Q^{(1)}_m,
\Qb^{(1)}_m$ on multi-soliton states.  This action
is defined by a comultiplication $\De$, which is determined
from the braiding relations of the currents $\CJ_m , \CJb_m$ with
the fields $K_m$ or $\Kb_m $, as in \IIxxiii .
As explained in \nlc , the contours defining the action of charges
on particle states can be closed if one considers the action of
chiral charges on anti-chiral soliton fields, and vice-versa.  From the
braiding relations \IIIxvii\braid\ one therefore deduces
\eqn\IIIxxx{\eqalign{
\De \( Q^{(1)}_m \) &= Q^{(1)}_m \ot 1 + \( R^{1,1/2} \)^n_m \ot Q^{(1)}_n
\cr
\noalign{\smallskip}
\De \( \Qb^{(1)}_m \) &= \Qb^{(1)}_m \ot 1 + \( \Rb^{1,1/2} \)^n_m \ot
\Qb^{(1)}_n
. \cr }}

Alternatively, these relations can be derived purely
quantum group theoretically. Since $Q_{-m}^{(1)}=\a_{Q_+}^{1-m} (Q_- )$
is given in terms of generators of \slqh, we can in principle
compute the coproduct explicitly. The computation is simplified
by the following observation \rBF. It follows from the general
form of the Hopf algebra \IIix\  that $\De ( Q^{(1)}_m )$
has the form \IIxxiii . The matrix $\Theta\in{\rm End}\, (V_1\ot V_{1/2})$
satisfies the intertwiner relation $\Theta\De=\De'\Theta$.
 It must therefore be proportional
to the $R$-matrix $R^{1,1/2}$. The proportionality constant
is determined by looking at the $n=m=1$ matrix element.

Finally, the charges $Q_+ , \Qb_- , T$ also have a well defined action
on states.  Since the Lorentz spin of these charges is zero, and
they generate the \slq algebra, one deduces that they have the following
representation on states
\eqn\IIIxxxi{
Q_+ = c \, \sigma_+ q^{\sigma_3 /2}, ~~~~~ \Qb_- = c\,
\sigma_- q^{\sigma_3 /2} , ~~~~~T= \sigma_3 , }
and $\De$ as in \IIix .

The 2-particle to 2-particle S-matrix as usual is an
operator $\S_{12} : V_1\ot V_2 \to V_2 \ot V_1$
where $V_{1,2}$ are the 2-dimensional vector spaces spanned by
$\ket{K_m (\th )}$.  It is required to be a solution to the
symmetry equations
\eqn\sym{
\S_{12} (\th )\ \De_{12} (a) = \De_{21} (a) \ \S_{12} , ~~~~
{\rm for} ~~  a=T,Q_+ , \Qb_- , Q_m^{(1)} ,
\Qb^{(1)}_m .}   It is a remarkable fact that a solution to
these symmetry equations can be obtained from the usual S-matrix of
SG solitons by a simple change of gradation.  This
is ultimately the explanation for why it is possible to
obtain the S-matrix for perturbations of the minimal models from the
S-matrix of the SG theory.  Let $\S^\prin$ be the minimal solution to
the equations \IIxxix , which is known to be the usual SG S-matrix.
Define a new S-matrix $\S^{\rm homo.}$:
\eqn\IIIxxxiii{
\sigma_{21}^{-1} \, \S^\prin \, \sigma_{12} = \S^{\rm homo.} , }
where
$\sigma_{12} = \sigma\ot \sigma$, and $\sigma= e^{\th\sigma_3 / 2\ga}$.
The conjugation by $\sigma$ has the effect of changing the representation
of $Q_+ , \Qb_-$ from the principal gradation \IIxxviii\ to the
homogeneous one.  Namely, let us refer to the representation
in \IIxxviii\ as $Q_+^\prin , \Qb_-^\prin$, and the representation
\IIIxxxi\ as $Q_+^{\rm homo.} , \Qb_-^{\rm homo.}$.  Then
\eqn\IIIxxxiv{
\sigma^{-1} \, Q_+^\prin \, \sigma = Q_+^{\rm homo.} , ~~~~~~~
\sigma^{-1} \, \Qb_-^\prin \, \sigma = \Qb_-^{\rm homo.} . }
In the present context the need to change gradation is simply
a consequence of the background charge in the energy-momentum tensor,
which alters the Lorentz spin of conserved charges.  That
$\S^{\rm homo.}$ is a solution to the symmetry equations
\sym\ was shown explicitly in \rBLii .

We emphasize that the solution $\S^{\rm homo.}$ to the
equations \sym\ does not represent a physically
meaningful S-matrix as it stands, in contrast to $\S^\prin$.
Indeed, only $\S^\prin$ can be made crossing symmetric.
The S-matrix $\S^{\rm homo.}$ is to be viewed as an intermediate
step in the construction of the S-matrix of the RSG
theory, which we describe in the next section.
The restriction of the SG theory relies on the
\slq invariance; to describe it one needs an S-matrix
for \slq covariant states and with covariant symmetries,
which is what $\S^{\rm homo.}$ represents.  Indeed
the construction in this section is very much in
the spirit of the work of Moore and Reshetikhin on
the \slq symmetry of CFT \ref\rMoR{\MR}.

\def\k#1#2{ {K_{j_{#1} j_{#2}}} }
\def\kb#1#2{ {\Kb_{j_{#1} j_{#2}}} }
\def\cj#1#2{ {\CJ_{j_{#1} j_{#2}}} }
\def\cjb#1#2{ {\CJb_{j_{#1} j_{#2}}} }

\bigskip\noindent
4b. {\it The Restriction}
\medskip

We will now use the results of the last subsection to
provide a derivation of the RSG S-matrix based on its
residual symmetries.  The primary mathematical technique
from the theory of quantum groups we will employ is the
so-called vertex-RSOS correspondence\ref\rpas{\Pasquierii}
\rkol , which finds its origin in classical lattice
statistical mechanics \rABF\rKyoto .

The restricted space of states is a subquotient of the
space $\cal H$ of states
of the sine-Gordon model. The restriction consists of two steps: the
first restriction is to solid-on-solid (SOS) states and the second
to restricted SOS (RSOS) states. The SOS space is by definition the
direct sum of sectors ${\cal H}_j$ of $\CU_q(sl(2))$ singular vectors
of topological charge $-2j$, ($j=0$, $1/2$, $1$, $\dots$),
\eqn\A{
{\cal H}_j=\{\psi\in {\cal H} : \bar Q_-\psi=0,\quad T\psi=-2j\psi
\}}
such that the representation of $\CU_q(sl(2))$ generated by these
singular vectors is irreducible of non-vanishing $q$-dimension.
 For the specific value of the
SG coupling $\hb$ \IIIiv ,
\eqn\limit{
q=-e^{-i\pi/p},
}
the latter condition  puts a limitation on the allowed values of the
quantum spin \ref\rpassal{\PasS}:
\eqn\BB{
j\le {p\over 2}-1.
}
This restriction of the allowed values of $j$ is implicit throughout
this section\foot{This limitation on the allowed spins is a more
restricted definition of SOS states than the usual one in
statistical mechanics}.

 It is convenient to represent the sector ${\cal H}_j$ as
\eqn\B{
{\cal H}_j={\rm Hom}_{\CU_q(sl(2))}(V_j,{\cal H})
}
where $V_j$ is the irreducible spin $j$ representation of
$\CU_q(sl(2))$. The isomorphism of \B\ with \A\ is the identification
of a singular vector with the subrepresentation it generates.

It is clear from the representation \B\ that we can also characterize
the space in terms of highest weight singular vectors (i.e.
vectors in the kernel of $Q_+$) of positive
topological charge, rather than lowest weight singular vectors \A.

Of course ${\cal H}$ is highly reducible as $\CU_q(sl(2))$-module
and we can restrict our attention to the invariant subspace spanned
by products of fundamental fields at given positions $x_1,\dots,
x_N$ (or rather smeared with a given test function) applied
to the vacuum:
\eqn\C{
K_{\alpha_1}(x_1)\cdots K_{\alpha_N}(x_N)|0>,
\qquad
\alpha_i=\pm 1/2.
}
The SOS states in the space \C\ are therefore given by
\eqn\D{
\oplus_{j=0}^{p/2-1}{\rm Hom}(V_j,V_{1/2}^{\otimes N}),
}
where Hom in this section denotes the space of $\CU_q(sl(2))$
module homomorphisms. The asymptotic states corresponding to
\C\ are
\eqn\CPRIME{
|K_{\alpha_1}(\theta_1)\cdots K_{\alpha_N}(\theta_N)>
}
and the asymptotic $N$-particle SOS states with given rapidities
are also described by the space \D\  . Both classes of states
\C\ and \CPRIME\ are supposed to form a dense subspace of the
SG Hilbert space. The space spanned by the vectors \CPRIME\ with
fixed rapidities is invariant under the full \slqh.
In the sequel we fix $\theta_1,\dots,\theta_N$ and consider
the vector space spanned by the vectors \CPRIME.

The Shapovalov bilinear form on $V_j$ with
$(v,v)=1$ for the highest weight vector $v$ and such that
$(E_1^\pm u,w)=(u,E_1^\mp w)$, $u,w\in V_j$, induces a bilinear
form on the space \D. The RSOS restriction of an $\CU_q(sl(2))$
module $M$ is the direct sum over $j\in\{0,1/2,1,\dots, p/2-1\}$
of the quotient ${\rm Hom'}(V_j,M)$
of ${\rm Hom}(V_j,M)$
by the kernel of (\ \ ,\ \ ). If $M$ is a tensor product
of irreducible modules, it can be described
explicitly as restricted path space (see below).

In going to the restricted model we have lost the $\CU_q(sl(2))$
symmetry, but it is to be expected that there is still a residual
symmetry coming from the rest of $\CU_q(\hat{sl(2)})$. In particular,
we expect that the spin 1 multiplet $Q_m^{(1)}$, when suitably
projected onto RSOS states, provides conserved charges for
the restricted model.

Before describing how this works, we have to make the notion
of conserved charge in a theory with superselection sectors ${\cal H}_j$
more precise. It is natural to define a conserved charge as an
operator $Q_{ij}$ from ${\cal H}_j$ to ${\cal H}_i$ such that
\eqn\CC{
S_iQ_{ij}=Q_{ij}S_j
.}
Here $S_i$ is the time evolution operator the $S$ matrix
 in the sector ${\cal H}_i$. Conserved charges can be multiplied
only if the sectors match, so they do not build an algebra but
rather the space of morphism of a category whose objects are
the sectors of the model.

Such charges $Q_{ij}$ can be easily constructed starting from the multiplet
$Q^{(1)}_m$. Fix for each pair of spin values $i$, $j$ compatible
with the fusion rules a \slq Clebsch-Gordan operator
\eqn\DD{
C_{ij}\in {\rm Hom}(V_i,V_1\otimes V_j).
}
In our $sl(2)$ example, this operator is unique up to proportionality.
Denote by $\CU_q(\hat{sl(2)})^{\rm ad}$ the space $\CU_q(\hat{sl(2)})$ with
 adjoint action of $\CU_q(sl(2))$.
View $Q^{(1)}_m$ as
\eqn\EE{
Q^{(1)}\in {\rm Hom}(V_1,\CU_q(\hat{sl(2)})^{\rm ad})
,}
so that $(Q^{(1)}\otimes 1)C_{ij}
\in {\rm Hom}(V_i,\CU_q(\hat{sl(2)})^{\rm ad}\otimes V_j)$
acts naturally as a linear
operator
\eqn\FF{
Q_{ij}:{\rm Hom'}(V_j,M)\to {\rm Hom'}(V_i,M),
}
for any $\CU_q(\hat{sl(2)})$-module $M$. For $\psi\in{\rm Hom}(V_j,M)$,
$Q_{ij}\psi$ is defined as the composition
\eqn\GG{
V_i\to \CU_q(\hat{sl(2)})\otimes V_j{\buildrel
1\otimes\psi\over\longrightarrow}
\CU_q(\hat{sl(2)})\otimes M\to M.
}
As $M$ we take $V_{1/2}^{\otimes N}$ or more generally
$V_{1/2}^{\otimes N}\otimes V_k$.
A parallel construction can be done for the operators $\Theta_m^n$,
viewed as an element of ${\rm End}_{\bf C} V\ot \CU_q \( \hat{sl(2)} \)$
with the intertwining
property
\eqn\HH{
\Theta\> \De(X)=\De'(X)\> \Theta, \qquad X\in \CU_q \( sl(2) \).
}
We get linear operators
\eqn\JJ
{
\Theta_{ij}^{kl}:{\rm Hom'}(V_j,V^{\otimes N}\otimes V_l)\to
                 {\rm Hom'}(V_i,V^{\otimes N}\otimes V_k).
}
The comultiplication rule \IIIxxx\
translates into an RSOS relation. For this we need the isomorphism
\ref\rFrKe{\FrKe}
\eqn\E{
{\rm Hom'}(V_j,V_{1/2}^{\otimes N}\otimes V_l)
\simeq \oplus_k
{\rm Hom'}(V_j,V_{1/2}^{\otimes N_1}\otimes V_k)\otimes
{\rm Hom'}(V_k,V_{1/2}^{\otimes N_2}\otimes V_l)
,}
valid if $N_1+N_2=N$. On the $k$th summand in this decomposition
we have
\eqn\F{
\ {}_NQ_{ij}=\ {}_{N_1}Q_{ij}\otimes \ {}_{N_2}1+
\sum_l \ {}_{N_1}\Theta_{ij}^{lk}
\otimes_{N_2}Q_{lk},}
where the dependence on the tensor power is displayed as a left subscript.
Iterating the decomposition \E\ we obtain
the restricted path space decomposition
\eqn\P{
{\rm Hom'}(V_j,V_{1/2}^{\otimes N}\otimes V_l)
=\oplus_{\rm paths}\otimes_{s=0}^{N-1}
{\rm Hom}(V_{j_{s+1}},V_{1/2}\otimes V_{j_s}).}
The direct sum is over all paths $j_0,\dots,j_s$
with $|j_{s+1}-j_s|=1/2$,
$j_0=j$, $j_N=l$, and $j_s\le p/2-1$.

A completely parallel construction can be done for the multiplet
$\bar Q^{(1)}$. In particular we have conserved charges $\bar Q_{ij}$
with RSOS ``comultiplication''
\eqn\G{
{}_N\bar Q_{ij}=\ {}_{N_1}\bar Q_{ij}
\otimes \ {}_{N_2}1+\sum_l \ {}_{N_1}\bar\Theta_{ij}^{lk}
\otimes_{N_2}\bar Q_{lk},}

In an abstract setting, what we have done is the following: we have
a Hopf algebra $A$ with a Hopf subalgebra $B$ and an $A$-module
$M$. The restricted model is given by sectors ${\cal H}_j=$
Hom${}_B(V_j, M)$, for some class of $B$-modules $\{V_j\}$.
 One then considers the category whose objects
are the modules $V_i$ and whose space of morphisms between $V_i$
and $V_j$ is Hom${}_B(V_i,A^{\rm ad}\otimes V_j)$. This category
acts by symmetries on the sectors
in the sense that there is a contravariant
functor to the category whose objects are the sectors and whose
morphisms are linear maps commuting with the time evolution.
In the above construction we have specialized this general setting
to the elements of Hom${}_B(V_i,A^{\rm ad}\otimes V_j)$ that come
from a homomorphism $Q^{(1)}$
from a particular $B$-module to $A^{\rm ad}$.

Let now see how the construction works in explicit terms.
{}From the fields $K_\al (x)$ and $\CJ_m (x)$ in
\IIIxxi\  and \IIIxix\
let us define new fields $\k  21 (x), j_1 = \j 2 \pm 1/2$
and $\cj 21 (x), \j 2 = \{ \j 1 , \j 1 \pm 1 \} $,
for $\j 2 , \j 1 \in \{0,1/2, 1, \ldots \} $ in the
following way:
\eqn\IIIxxv{\eqalign{
\k n{n-1} (x_n ) &\k {n-1}{n-2} (x_{n-1} ) \cdots
\k 10 (x_1 ) \ket{0}  \cr
&=
\sum_{m_i }
\k n{n-1} (x_n )^{-\j n}_{m_{n-1}}
\k {n-1}{n-2} (x_{n-1} )^{m_{n-1}}_{m_{n-2}} \cdots
\k 10 (x_1 )^{m_1}_{m_0} \ket{0} \cr
\cj n{n-1} (x_n ) &\cj {n-1}{n-2} (x_{n-1} ) \cdots
\cj 10 (x_1 ) \ket{0} \cr &=
\sum_{m_i}
\cj n{n-1} (x_n )^{-j_n}_{m_{n-1}}
\cj {n-1}{n-2} (x_{n-1} )^{m_{n-1}}_{m_{n-2}} \cdots
\cj 10 (x_1 )^{m_1}_{m_0} \ket{0} \cr
}}
where $j_0 = m_0 = 0$,
\eqn\IIIxxvi{\eqalign{
\k 21 (x)^{m_2}_{m_1} &= \sum_\al \cg {\j 2}{1/2}{\j 1}{m_2}\al{m_1}
\ K_\al (x) \cr
\cj 21 (x)^{m_2}_{m_1} &= \sum_m \cg {\j 2}{1}{\j 1}{m_2}m{m_1}
\ \CJ_m (x) . \cr
}}
The fields $\kb 21 (x)$ and $\cjb 21 (x)$ are defined similarly
from $\Kb_\al (x)$ and $\CJb_m (x)$.
More generally, one may define fields $\phi^{(j)}_{\j 2 \j 1} (x)$
and $\phib^{(j)}_{j_2 j_1 } (x)$ from $\phi^{(j)}_m , \phib^{(j)}_m$.

The product of fields in \IIIxxv\ is characterized as being
a lowest weight vector in a \slq representation and gives an
explicit basis of the path space \P.  Note that
$K_{\half 0} (0) \ket 0 = \phi^{(1/2)}_{-1/2} (0) \ket 0 $ and
\eqn\IIIxxxvii{
K_{j \half} (x) K_{\half 0} (0) \ket 0 = \sum_m K_{j\half}
(x)^{-j}_m \phi^{(1/2)}_m (0)
\ket 0 . }
The state $\phi^{(1/2)}_{-1/2} (0) \ket 0 \equiv \ket{1/2} ^\cft $
is a state of the CFT, namely the one corresponding to the field
$\ph 21$.  More generally define
\eqn\IIIxxxviib{
\phi^{(j)}_m (0) \ket 0 \equiv \ket j _m^\cft . }
The states $\ket j _{-j}^\cft $ are the primary states
$\ket {(2j+1,1)} = \ph {2j+1}1 (0) \ket 0 $ of the minimal CFT.
(These states are not to be confused with the massive particle-like
asymptotic states of the perturbed theory.)  The generalization
of \IIIxxxvii\ is
\eqn\IIIxxxviii{
\k 21 (x) \ket {j_1} = \sum_{m_1} \k 21 (x)^{-j_2}_{m_1}
\ket {j_1} _{m_1}^\cft . }
The fields $\k 21 (x)$ expressed in the form \IIIxxxviii\
are nothing other that the intertwiners (chiral vertex operators)
for the minimal model field $\ph 21 (x)$\rGoS\rFW . Similar
arguments apply to the fields $\cj 21$.  More precisely
we have the identification
\eqn\IIIxxxix{\eqalign{
\k 21 (x) &\sim \phi^{(2,1)}_{(2\j 2 +1,1)(2j_1+1,1)} (x) \cr
\cj 21 (x) &\sim \phi^{(3,1)}_{(2\j 2 +1,1)(2j_1+1,1)} (x) , \cr
}}
where the fields on the RHS are the chiral vertices for $\ph 21 ,\ph 31$.
The restriction of allowed spins \limit\ is of course consistent with the
limitation of the $(n,m)$ indices of the primary fields $\ph nm$.

We will need the braiding relations for the fields in \IIIxxv .
By using \IIIxxv , the braiding relations \IIIxvii\braid , and the
identity \rkol :
\eqn\eid{\eqalign{
\sum_{m_2 ,\al_1 , \al_2 }
&\cg {j_3}{j'}{j_2}{m_3}{\al_1}{m_2}
\cg {j_2}j{j_1}{m_2}{\al_2}{m_1}
\( R^{j'j} \)^{\al_1 '\al_2 '}_{\al_1 \al_2} \cr
&= \sum_{j_4 ,m_4}
\( \TH^{j'j} \)^{\j 4\j 1}_{\j 3 \j 2}
\cg {j_3}{j}{j_4}{m_3}{\al_2 ' }{m_4}
\cg {j_4}{j'}{j_1}{m_4}{\al_1 '}{m_1} , \cr }}
\smallskip
\eqn\eth{ \( \TH^{j'j} \)^{j_4 j_1 }_{j_3 j_2}
= (-)^{\j 2 + \j 4 -\j 1 -\j 3 }
q^{C_{j_1} + C_{j_3} - C_{j_2} -C_{j_4} }
\sj {j'}{j_3}{j_2}j{j_1}{j_4}
}
where $C_j = j(j+1)$, and $\{ * \}_q$ are q-6j symbols, one
can show that the fields satisfy the non-abelian RSOS braiding
relations
\eqn\IIIxxxxi{\eqalign{
\phi^{(j')}_{j_3 j_2} (x) \ \phi^{(j)}_{j_2 j_1} (y)
&= \sum_{j_4} \( \TH^{j'j} \)^{j_4 j_1}_{j_3 j_2} \
\phi^{(j)}_{j_3 j_4} (y) \ \phi^{(j')}_{j_4 j_1} (x) ~~~~~x<y \cr
\phib^{(j')}_{j_3 j_2} (x) \ \phib^{(j)}_{j_2 j_1} (y)
&= \sum_{j_4} \( \THb^{j'j} \)^{j_4 j_1}_{j_3 j_2} \
\phib^{(j)}_{j_3 j_4} (y) \ \phib^{(j')}_{j_4 j_1} (x) ~~~~~x<y \cr
\phib^{(j')}_{j_3 j_2} (x) \ \phi^{(j)}_{j_2 j_1} (y)
&= \sum_{j_4} \( \TH^{j'j} \)^{j_4 j_1}_{j_3 j_2} \
\phi^{(j)}_{j_3 j_4} (y) \ \phib^{(j')}_{j_4 j_1} (x) ~~~~~\forall\ x,y \cr
}}
where
\eqn\ethb{ \( \THb^{j'j} \)^{j_4 j_1 }_{j_3 j_2}
= (-)^{j_1 + j_3 -j_2 -j_4 }
q^{C_{j_2} + C_{j_4} - C_{j_1} -C_{j_3} }
\sj {j'}{j_3}{j_2}{j}{j_1}{j_4} .
}
The 1-1 q-6j symbols are presented in the appendix.
Smirnov has also considered fields with such braiding relations
in the present context\ref\rsmirii{\Smirii} .

Just as the asymptotic states $\ket{K_\al (\th )}$ are created by
the fields $K_\al (x)$, $\Kb_\al (x)$, we define states created
by the fields $\k 21 (x)$, $ \kb 21 (x)$:
\eqn\IIIxxxxiii{\eqalign{
| \k n{n-1} (\th_n ) &\k {n-1}{n-2} (\th_{n-1} ) \cdots \k 10 (\th_1 )
\rangle\cr
&= \sum_{m_i}
| \k n{n-1} (\th_n ) \rangle^{-j_n}_{m_{n-1}}
|\k {n-1}{n-2} (\th_{n-1} ) \rangle^{m_{n-1}}_{m_{n-2}}
\cdots |\k 10 (\th_1 ) \rangle^{m_1}_{m_0}  \cr}}
($j_0 =m_0 =0$) where
\eqn\IIIxxxxiv{
|\k 21 (\th ) \rangle^{m_2}_{m_1} = \sum_\al
\cg {j_2}{1/2}{j_1}{m_2}\al{m_1} \ket{ K_\al (\th )}
. }
The fields $\k 21 , \kb 21$ are not the unique fields that
create the above states; the product of these fields with any
 field in the vacuum sector can also create them.

\def\q#1#2{ {Q_{j_{#1}  j_{#2}}} }
\def\qb#1#2{ {\Qb_{j_{#1}  j_{#2}}} }

 Since the currents $\CJ_m , \CJb_m$ are conserved in the
 perturbed theory, so are $\cj 21 , \cjb 21$, which define
 conserved charges $\q 21 , \qb 21$, with Lorentz spin
 given in  \IIIxxb .
These charges have expressions in terms of $Q^{(1)}_m ,
\Qb^{(1)}_m$ obtained by integrating \IIIxxvi .
Indeed one can show in an  intrinsic  minimal model description \rBLii\
(i.e. without reference to the Feigin-Fuchs construction) that the
fields  $\ph 31 , \phb 31$ define conserved currents for the action
\Iib , i.e.
\eqn\eintt{
\d_\zb \ph 31 = \lambda C \ \d_z \( \ph 33 \phb 13 \) ,}
where $C$ is a structure constant, which gives the conserved charges
\eqn\echarge{\eqalign{
Q&=  \int \frac{dz}{2\pi i}  \> \ph 31 + \int \frac{d\zb}{2\pi i}
\( \lambda C
\ph 33 \phb 13 \) \cr
\Qb & =  \int \frac{d\zb}{2\pi i}
\> \phb 31 + \int \frac{dz}{2\pi i}  \( \lambda C \phb 33 \ph 13 \) . \cr
}}
The charges $\q 21 , \qb 21$ represent the above charges in specific
superselection sectors.

In the CFT the current $\phi^{(3,1)}$ plays a special role:  for each $p$
it generates a fractional spin chiral algebra with a series of minimal
unitary representations, where the usual $p-th$ minimal model is the lowest
member of this series\ref\rcoset{\cosetFF} .  For recent results regarding
this chiral algebra see \ref\rAGT{\AGT}.

We now derive the implications of the conserved charges
$\q 21 ,\qb 21$ for the S-matrix of  the states \IIIxxxxiii .
Consider the action of the charges on a 1-kink state.  One has
\eqn\IIIxxxxvi{
\( \q 32 \ket{ \k 21 (\th )} \)^{m_3}_{m_1} =
\sum_{m_2 , m,\al}
\cg {j_3}1{j_2}{m_3}m{m_2}
\cg {j_2}{1/2}{j_1}{m_2}\al{m_1} \ Q^{(1)}_m \, \ket{K_\al (\th )} . }
Using the matrix elements \IIIxxix , and the identity\rkol
\eqn\IIIxxxxvii{\eqalign{
\sum_{m,m_2 ,\al}
\cg {j_3}1{j_2}{m_3}m{m_2}
&\cg {j_2}{1/2}{j_1}{m_2}{\al}{m_1}
\cg 1{1/2}{1/2}m\be\al = \cr
&\sj {j_3}1{j_2}{1/2}{j_1}{1/2}
\cg {j_3}{1/2}{j_1}{m_3}{\be}{m_1} , \cr}}
one finds
\eqn\IIIxxxxviii{
\q 32 \ket{ \k 21 (\th ) } =
e^{2\th /p}
\sj {j_3}1{j_2}{1/2}{j_1}{1/2}
\ket{ \k 3 1 (\th ) } . }
Similarly
\eqn\IIIxxxxviiib{
\qb 3 2 \ket{ \k 2 1 (\th ) } =
e^{-2\th /p}
\sj {j_3}1{j_2}{1/2}{j_1}{1/2}
\ket{ \k{3}{1} (\th ) } . }

This formula is the expression in the path space basis of the action
of $Q_{j_3j_2}$ on Hom${}_{\CU_q(sl(2))}(V_{j_2},V_{1/2}\otimes V_{j_1})$.

Using the comultiplication formula \F , and the path space decomposition
\P, one can compute the action of $Q$ on arbitrary path spaces.
One has for instance
\eqn\IIIxxxxx{\eqalign{
\q 32 \ket{ \k 21 (\th_2 ) \k 10 (\th_1 )}
&= \( \q 32 \ket{ \k 21 (\th_2 ) } \) \ket {\k 10 (\th_1 )}
\cr &~~~+ \sum_{j_4} \(  \hat{\TH}^{j_4 j_1}_{j_3 j_2}
\ket{ \k 21 (\th_2 )} \) \( \q 41 \ket{ \k 10 (\th_ 1 ) } \)
,\cr }}
where
\eqn\IIIxxxxxi{
 \hat{\TH}^{j_4 j_1}_{j_3 j_2} \ket{\k 21 (\th ) }
= \( \TH^{1,1/2} \)^{j_4 j_1}_{j_3 j_2} \ket {\k 34 (\th )} . }

The RSG S-matrix $S^{j_1 j_0}_{j_2 j_3} (\th )$ was conjectured to describe
the 2-kink to 2-kink scattering process
\eqn\IIIxxxxxii{
\ket{ \k 21 (\th_2 ) \k 10 (\th_1 ) } \to
\ket{K_{j_2 j_3}  (\th_1 ) K_{j_3 j_0}  (\th_2 ) } }
($j_0$ is no longer  necessarily $0$.)  As usual this S-matrix is
required to commute with the conserved charges $\q 21 , \qb 21$,
whose action is given above.  Starting from the conjectured  RSG S-matrix,
it was shown in \rBLii\ that this S-matrix does indeed possess these
on-shell symmetries.  Our analysis thus provides a derivation of
the RSG S-matrix if it happens to be the minimal\foot{
The solutions to the symmetry equations can only be unique
up to overall scalar functions of rapidity.  For minimal solutions,
these overall factors are the minimal ones that are
required for crossing symmetry and unitarity.} solution to the
symmetry equations.  This is likely to be the case since the symmetries
we have constructed are inherited from the \slqh symmetry of the
SG theory, and this latter symmetry does provide a unique minimal
solution.  For completeness we present the solution
\eqn\smat{\eqalign{
S^{j_1 j_0}_{j_2 j_3} (\th ) &= \frac{u(\th )}{2\pi i}
\( \frac{ [2j_1 +1]_q [2j_3 +1]_q}{[2j_2 +1]_q [2j_0 +1]_q }
\)^{-\th/2\pi i} \cr
& \times \left\{ \sinh (\th /p ) \delta_{j_2 j_0}
\( \frac{[2j_1 + 1]_q [2j_3 +1]_q }{[2j_2 +1]_q [2j_0 + 1]_q } \)^{1/2}
+ \sinh \( \frac{i\pi - \th}{p} \) \delta_{j_1 j_3} \right\} , \cr } }
where $u(\th )$ is a scalar function defined in \rBLii .

\def\T#1#2#3#4#5#6{{ \( \Th^{#1#2} \)^{j_{#6} j_{#3}}_{j_{#5} j_{#4} } }}
\def\Tb#1#2#3#4#5#6{{ \( \Thb^{#1#2} \)^{j_{#6} j_{#3}}_{j_{#5} j_{#4} } }}
\def\That#1#2#3#4{{  \hat{\Th}^{j_{#4}
j_{#1}}_{j_{#3} j_{#2} } }}


\bigskip
\noindent
4c. {\it The Restricted Quantum Affine Symmetry}

\bigskip
As explained in the last sub-section, after restriction the
remnant of the quantum affine symmetry is the conserved charges
$Q, \Qb$ defined in \echarge .   Evidently, these charges no longer
satisfy the \slqh relations.
The residual symmetries obey new relations which we now characterize.

The relations satisfied by the
charges should be characterized by braided commutators,
as explained for the \slqh case in section 2b.  From the braiding
relations \IIIxxxxi , one finds the RSOS analog of \IIxx , \IIxxi :
\eqna\resi
$$\eqalignno{
\a_{\q 32} \( \phi^{(j)}_{j_2 j_1 } (y) \) &=
\q 32 \> \phi^{(j)}_{j_2 j_1} (y)
-\sum_{j_4} \T 1j1234 \> \phi^{(j)}_{j_3 j_4 } (y) \> \q 41
&\resi{a} \cr
\a_{\q 32} \( \phib^{(j)}_{j_2 j_1 } (y) \) &=
\q 32 \> \phib^{(j)}_{j_2 j_1} (y)
-\sum_{j_4} \Tb 1j1234 \> \phib^{(j)}_{j_3 j_4 } (y) \> \q 41
, &\resi{b} \cr}$$
and similarly for
$\a_\Qb \( \phib^{(j)} (y) \)$ and
$\a_\Qb \( \phi^{(j)} (y) \)$.
We will also need the adjoint action on a product of fields, which is
given by the comultiplication in \F , \G :
\eqn\resii{\eqalign{
\a_{\q 32} \( \phi^{(j)}_{j_2 j_1} (x) \> \phi^{(j')}_{j_1 j_0} (y) \)
= &\a_{\q 32}  \( \phi^{(j)}_{j_2 j_1} (x) \) \>
\phi^{(j')}_{j_1 j_0} (y) \cr
&+ \sum_{j_4} \That 1234 \( \phi^{(j)}_{j_2 j_1} (x) \)
\> \a_{\q 41} \( \phi^{(j')}_{j_1 j_0} (y) \) , \cr } }
where
\eqn\resiii{
\That 1234 \( \phi^{(j)}_{j_2 j_1} (x) \)
= \T 1j1234 \> \phi^{(j)}_{j_3 j_4} (x) . }

We first consider the analog of the \slqh relations \IIx .
As shown in \nlc , one can always close the contour in the
adjoint action of chiral on anti-chiral conserved charges;
using this result one can show that
\eqn\resiv{
\a_{\q 32} \( \qb 21 \) =
\q 32 \qb 21 - \sum_{j_4} \Tb 111234 \> \qb 34 \q 41 =
\tilde{T}_{j_3 j_1}  }
where $\tilde{T}_{j_3 j_1}$ is an intertwiner for the
topological charge of the conserved current
$\frac{\lambda C}{2\pi i} \epsilon^{\mu\nu} \d_\nu \Phi^{(3,3)} (x)$:
\eqn\resv{
\tilde{T} = \frac{\lambda C}{2\pi i} \int dx \> \d_x \Phi^{(3,3)} (x) . }

\def\TT#1#2#3#4{{ \Th^{j_{#4} j_{#1}}_{j_{#3} j_{#2} } }}
\def\TTT#1#2#3#4{{ \Th^{{#4} {#1}}_{{#3} {#2} } }}

A more interesting question is whether there are additional relations
of the Serre type.
  The way to answer this question was outlined in section 2c.
Namely consider $\a_Q^n \( \CJ^\mu (y) \) $, where
$\CJ^\mu (y) = (\phi^{(3,1)} , \lambda C \phi^{(3,3)} \phib^{(1,3)} )$ is
the non-local current for the charge $Q$.             Displaying sectors,
one considers
\eqn\resvi{
\a_{\q {n+1}n } \cdots \a_{\q 32} \( \a_{\q 21} \( \CJ_{j_1 j_0} (y) \) \) . }
In the abelian case the condition for closure of the contours was
\servi .  Using similar arguments, it is not difficult to see that
the condition for closure can now be formulated as follows.
For the remainder of this section let $\Th^{j_4 j_1}_{j_3 j_2}
\equiv \T 111234 $.
Let $v(j_n , \ldots , j_1 )$, where
$j_{n+1} = j_n , j_n \pm 1$, span a vector space.
Define a matrix $M_{(j_{n+1} ,j_0 )}$ which acts on this vector space,
with the matrix elements:
\eqn\resvii{
\( M_{(j_{n+1} , j_0 )} \)^{j'_n \cdots j'_1}_{j_n \cdots j_1}
\equiv \sum_k  \Th^{j'_n j'_{n-1} }_{j_{n+1} j_n } \cdots
\Th^{j'_3 j'_2}_{j_4 j_3} \> \Th^{j'_2 j'_1}_{j_3 j_2 }
\> \Th^{j'_1 j_0 }_{j_2 k}
\>
\Th^{k j_0}_{j_2 j_1 }  . }
The spins $j_{n+1} , j_0$ correspond to the initial and
final sectors.  $M_{(j_{n+1} , j_0 )}$ represents the product
of braiding factors which is the non-abelian generalization of
the first summand of the term in braces in \serv , and
is represented graphically in figure 3.
$$
\matrix{
j_{n+1}       & j'_n &   & &j'_2   &   j'_1     &j_0          \cr
       &      &                                 \cr
       &      &                                 \cr
       &      &                                 \cr
       &      &                                 \cr
       &      &                                 \cr
       &      &      &   &   &    k   &          \cr
       &      &                                 \cr
j_{n+1}&   &      &   &j_2   &   j_1  & j_0      \cr
}
$$
\vskip 10pt
\centerline{\eightrm Fig. 3. The product of braiding
matrices representing the matrix
$M_{(j_{n+1} ,j_0 )}$.}
\vskip 10pt

\noindent
The braiding matrices $\TT 0123$ in the definition of $M$ are of
course subject to consistency with the fusion rules, i.e.
$j_{i+1}$ must appear in the fusion $j_i \times 1$, for
$i=0,1,2,3; j_4 \equiv j_0$.

Let $\sum_{j_1 , \ldots , j_n } c^{j_n \cdots j_1 }
\> v(j_n , \ldots , j_1 )$ be an eigenvector of $M$ with
eigenvalue $1$.  Then the contours may be closed in the
expression
\eqn\resviib{
\sum_{j_1 , \ldots , j_n } c^{j_n \cdots j_1} ~
\a_{\q {n+1}n } \cdots \a_{\q 32} \( \a_{\q 21} \( \CJ_{j_1 j_0} (y) \) \)  }
to give a well-defined operator.

\def\ha{{\frac{1}{2}}}

The above closure condition is difficult to implement in
practice, so we only present a few interesting examples.
It is clear from the explicit values of the braiding matrices
$\TH$ that the closure condition is not fulfilled for a
fixed $n$ for all $p$.  The simplest example occurs at $p=4$.
The relevant fusion rules are $0\times 1 = 1, 1\times 1 = 0,
1/2 \times 1 = 1/2$.  Using the q-6j symbols in the appendix one
finds for $q=-\exp (-i\pi /4 )$ that
\eqn\resviii{
\TTT 0101 = \TTT 1010 = \TTT \ha\ha\ha\ha = -1, ~~~~~(p=4), }
thus the braiding is effectively abelian.  Using the fusion rules one
finds three eigenvectors of the matrix $M$ for $n=1$:
$v(0,1,0), v(1,0,1), v(1/2,1/2,1/2)$, and by \resviii\ these have
eigenvalue $1$.  Thus the contour in \resvi\ can be closed for $n=1$.
Since $Q$ and $\CJ_{j_1 j_0} (y)$ have scaling dimension $1/2$ and
$3/2$ respectively, and $1\times 1 = 0$, $\a_Q \( \CJ \)$ must be a dimension
$2$  operator in the identity sector.  Thus we conclude that
\eqn\resix{
\a_{\q 21} \( \CJ_{j_1 j_0} (y) \) = T_{zz} (y),~~~~~
\a_{\qb 21} \( \CJb_{j_1 j_0} (y) \) = T_{\zb\zb} (y) }
\eqn\resx{
\Rightarrow  Q^2 = P, ~~~~~~~~\Qb^2 = \bar{P} , }
where $P, \bar{P} $ are the light-cone momentum operators.
This resulting supersymmetry is not unexpected since the $p=4$
minimal CFT is known to be supersymmetric; the $\Phi^{(1,3)}$
perturbation preserves the supersymmetry, as first noticed by
A. Zamolodchikov\rz .  One can check that the on-shell representation
\IIIxxxxviii\ indeed satisfies \resx ,
since $P=e^\th , \bar{P} = e^{-\th } $.

\def\q#1#2{{ Q_{#1#2} }}

A more  non-trivial example occurs at $p=6$.  Now the fusion rules with
spin-1 are $0\times 1 = 1, 1\times 1 = 0+1+2$, $2\times 1 = 1$.   Using
these we found three eigenvectors of $M$ for $n=2$ that are spanned by
single vectors: $v(0,1,1,0), v(2,1,1,2)$, and $v(2,1,1,0)$. Moreover,
for $q=-\exp (-i\pi /6 )$ these eigenvectors have eigenvalue $1$:
\eqn\resxi{
\TTT 1121 \TTT 0111 \TTT 0111 = \TTT 1101 \TTT 0111 \TTT 0111
= \TTT 1112 \TTT 2111 \TTT 2111 = 1. }
Scaling arguments indicate that the operator in \resviib\ for $n=2, p=6$
is again of dimension $2$.  Comparing the initial and final sectors for
the eigenvectors $v(0,1,1,0)$ and $v(2,1,1,2)$, one concludes from the
fusion rules that the operator in \resviib\ for these two cases is
necessarily proportional to the energy-momentum tensor.  Thus
\eqn\resxii{
\a_{\q 01} \( \a_{\q 11} \( Q_{10} \) \)
=
\a_{\q 21} \( \a_{\q 11} \( Q_{12} \) \)
= \tilde{c} \> P, }
for some constant $\tilde{c}$,
and similarly for $\a_\Qb^2 (\Qb )$.
The above relations may be written out explicitly using the integrated
versions of \resi{}  and \resii .  We illustrate the first relation
in \resxii :
\eqn\resxiii{\eqalign{
\sum_{j,k,l} \q 01 \q 11 \q 10 &- \TTT 011j \> \q 01 \q 1j \q j0
- \TTT 110k \TTT 01kl \> \q 0k \q kl \q l0  \cr
&+ \TTT 011j \TTT j10k \TTT 0jkl \> \q 0k \q kl \q l0
= \tilde{c} \> P . \cr}}

\bigskip
\noindent
4d. {\it The Ultraviolet Limit of the Restricted Model}
\medskip

The ultraviolet limit of the sine Gordon model for
$\hat \beta < \sqrt{2}$ is given by a free boson field.
Indeed the cosine perturbation is relevant in this
range and is therefore invisible at short distances.
The scaling dimensions of the exponentials are modified
by the additional term in the energy-momentum tensor
\IIIi\  of the restricted SG model, in such a way that
one of the exponential becomes marginal. The ultraviolet
limit is then the Liouville model with imaginary coupling
\eqn\UVi{
 S\ =\ \inv{4\pi}\int d^2z\ \d_z\Phi \d_{\zb}\Phi \
+ \frac{\la}{\pi }\ \int d^2z\ :\exp\(i\hb \Phi\): ~~.
}
This theory is known to be conformally invariant. For the
discrete series of values of the coupling \IIIiv\ it is
believed to be equivalent to the unitary minimal models when
suitably restricted \ref\rCThorn{\CThorn}
\ref\rGerNev{J.-L. Gervais and
A. Neveu, Nucl. Phys. B 238 (1984) 125}
\ref\rGeri{\Geri} .

Let us compare the restriction defined in this paper with
the restriction appearing in the free field representation
of minimal models. We discuss this for holomorphic chiral
fields. A similar discussion applies to antiholomorphic
fields. The primary fields $\ph {2j+1}1$ of  the CFT are
identified with the operators \IIIviii . Note that
since  $V_{2j+1,1}$ is local with respect to the  Liouville
perturbation and has regular operator product expansion
with it, the corresponding  primary fields is holomorphic.
However, the screening operator, which is identified
with the current $\exp (-2i \phi/\hb)$,
 has a double pole singularity in its operator product with
the perturbation. Hence it is conserved but not holomorphic. Thus
this construction gives a slightly different but equivalent
description of  minimal models.

Correlation functions of fields \IIIviii\ in the conformal limit
form a vector space isomorphic to a tensor product
of \slq representations. They
do not in general obey the conformal Ward identities. The
violation is due to boundary terms in the integration over
the position of the screening operators. It was remarked in
\ref\Trieste{G. Felder, Proceedings of the ICTP conference
on ``New developments in conformal field
theory'', S. Ranjbar-Daemi and J.-B. Zuber (eds.),
World Scientific (1990)}\rGoS\rFW that the
linear combinations of such correlators that obey the
conformal Ward identities are in fact identified
with \slq singular vectors in the tensor product. Indeed, the
step operator is identified topologically with a boundary
operator acting on relative homology groups \rFW, so that
absolute cycles are in its kernel. This gives part of the SOS
restriction, namely the restriction to singular vectors.

The rest of the RSOS restriction in the UV limit is
understood in terms of the Fock space cohomology
defined in \rfeld. The unitary truncation of
the Fock space of (topological) charge $1-m$
is the cohomology group Ker $Q_+^m\ /$ Im $Q_+^{p-m}$.
It was noted by Pasquier and Saleur \rpassal that this
cohomology on the space of singular vectors in the
 tensor product of spin $1/2$ representations
gives precisely the restricted path space described above.
 In other words, we see that the discrete series
of restricted SG models are given in their
UV limit by minimal models in their free field
representation. The restriction in the UV limit
consists of two parts: the topological part ensures
the validity of conformal Ward identities. The
RSOS restriction is equivalent to the Fock space
cohomology, which selects irreducible unitary
Virasoro representations in the Fock space.

\bigskip
\noindent
4e. {\it Remarks on the Degenerate Vacuum Structure}
\medskip

For $p=4,6$ A. Zamolodchikov has argued from a microscopic
analysis that the $\Ph 13$ perturbation of the minimal conformal
models has 3 and 5 degenerate vacua respectively, and
interpreted the kinks $\ket{ \k 21 (\th )} $ as interpolating
the two vacua labeled $j_1$ and $j_2$.
The 3-fold vacuum degeneracy was checked numerically for $p=4$
in \ref\rLMC{\LMC} , by using the
truncated conformal space approach.
The qualitative  aspects of
this picture was extended to all p in \rBLii\ by examining the
topological charge of the vertex operators $V_{nm}$.  Namely,
since the topological charge of the operators $V_{n1}$ is
$1-n$, and $1\leq n\leq p-1$, the minimum topological charge is
$2-p$.  On the other hand classical field configurations of
topological charge $n$ satisfy $\bh \( \Phi (\infty ) - \Phi (-\infty )
\) = 2\pi n$, and thus connect the wells in the $\cos (\bh \Phi )$
SG potential.
Thus the wells are effectively $p-1$ in number.  The problem with
the latter argument is that it does not allow an intrinsic description
of these degenerate vacua in terms of the usual properties of the
minimal conformal models.  Recently, based on the thermodynamic
Bethe-ansatz equations of Al. Zamolodchikov, Klassen and Melzer
conjectured that the lowest excited states in finite volume
are the local states associated with $\Ph nn (0) \ket{0}$.  In infinite
volume these states become degenerate with the vacuum, and there
are precisely $p-1$ in number.
The 1-kink states we have described in the previous section are associated
with the {\it chiral} CFT states $\ph 21 (0) \ket{0}$, and these are to
be interpreted as the first excited states above the vacua in infinite
volume.

We now offer some justification for the above identification of
the vacua in the approach we are developing.  To do this one must
identify a suitable order parameter.  For the SG theory, one may take
the SG field $\Phi$ itself.  The infinitely many degenerate vacua
of the $\cos (\bh \Phi )$ potential occur at $\Phi = 2\pi n / \bh ,
n \in \Z$.  However a different choice of order parameter is more
suitable for generalization.  In \nlc\ it was shown that the
q-affine charges satisfy
\eqn\ord{
Q_{\pm}\ \bQ_{\mp} - q^{-2}\ \bQ_{\mp}\ Q_{\pm}\ =\
\frac{\la}{2\pi i}
\( \frac{\bh^2}{\bh^2 -2} \)^2
\ \int_t dx\ \d_x
\exp\(\pm i\(\frac{2}{\hb}-\hb\)\Phi(x,t)\) . }
The RHS of the above equation may be considered as a generalized
topological charge (which is actually a function of the usual
topological charge).  So let us define the order parameters of the
SG theory as $\CO^{SG}_\pm = \exp \(\pm i \( \frac{2}\bh - \bh \)
\Phi (x) \)$.  The fields  $\CO^{SG}_\pm$ are  local scalar fields
which define topological currents $\tilde{J}^\mu_\pm (x) =
\ep^{\mu\nu} \d_\nu \CO^{SG}_\pm (x)$.  This choice of order
parameter is in a sense more fundamental than $\Phi$ since its
topological charge
appears in the symmetry algebra.
Thus, if the symmetry algebra has a well-defined representation
on the states, this implies that $\CO^{SG}_\pm (\infty )
-\CO^{SG}_\pm (-\infty )$ is well-defined on these states.
If we further suppose that the topological particle spectrum
has the `kink' property of connecting degenerate vacua, then
$\CO^{SG}_\pm$ must probe these vacua.
At the location of the
degenerate vacua $\Phi = 2\pi n /\bh$, the order parameters take
the values $\CO^{SG}_\pm  = \exp (\pm 4\pi i n / \bh^2 )$.
For irrational values of the coupling constant $\bh$, all of the
vacua are distinguished by this order parameter.

Let $Q, \Qb$ be the residual quantum affine charges in
\echarge .  For the perturbed minimal CFT the relation \ord\
becomes  \resiv .
This suggests that an appropriate order
parameter for the RSG theory is $\Ph 33 (x)$.  Recall that
$\Ph 33 = : (\Ph 22 )^2 :$, thus
$\Ph 22$ is superior as an order parameter, since it is more
refined.  In the
Landau-Ginzburg description of the p-th minimal model\ref\rzlg{
\Zamovi},
the Landau-Ginzburg scalar field was identified precisely
with $\Ph 22$, and the Landau-Ginzburg potential is of degree
$2(p-1)$ and can support $p-1$ degenerate vacua.  Thus our
identification of the order parameter is entirely consistent
with the Landau-Ginzburg picture.   Considering the
Feigin-Fuchs representation of the field $\Ph 22$, one sees
that it takes the values $(1, -q , (-q)^2 , ....)$ at the
wells of the SG potential.  Since
$: (\Ph 22 )^k : = \Ph {1+k}{1+k}$, one can thus
associate the degenerate vacua with the states
$\Ph nn (0)\ket{0}$.  However the precise connection between
the minima of the SG cosine potential and the minima of the
Landau-Ginzburg potential has not been clarified and is
an interesting problem.

 Note that in finite volume (on the cylinder) one should not expect that
the degeneracy of the classical vacua of the LG potential
corresponds to degenerate ground states of the quantum
theory. Indeed tunneling between the minima of the
potentials induces a splitting of the energy levels, and
this explains why the states $\Ph nn (0)|0\rangle$ have
non-degenerate energies in the CFT. The
true ground state is described by a wave function on
constant field configurations without nodes. The first
excited states are then described by wave functions with
nodes, and are obtained from the ground state by applying
suitable polynomials in the order parameter at imaginary
time $-\infty$. These polynomials are the normal ordered
powers of the order parameter $\Ph {1+k}{1+k}(0)$ in the
conformal limit, and the coefficients of these polynomials
are precisely chosen so that the corresponding states are
eigenvectors of the hamiltonian.   In the infinite volume
limit tunneling is suppressed, the splitting disappears
and in the low temperature phase degenerate vacua appear.

\newsec{Conclusions}

\medskip

The techniques we have described in this work should extend
to the other models listed in the introduction, though the
details have not been worked out.  In particular, the formulation
of the fractional supersymmetric sine-Gordon models and their
restrictions to the perturbed cosets
$su(2)_k \otimes su(2)_l /su(2)_{k+l}$ lends itself straightforwardly
to the above treatment.  Again, our characterization of superselection
sectors provides a qualitative understanding of the spectrum.
For example,  for the current-current perturbation of the Wess-Zumino-Witten
theories at level-$k$, the massive kink spectrum $K^\pm_{j_2 j_1}$ is
created by the chiral intertwiners $\phi^{(1/2)}_{j_2 j_1}$ for the
spin-1/2 primary field of the CFT.

The new frontier of this subject is the determination of the
off-shell properties of the models, namely the form factors and
correlation functions, from symmetry principles.  Since
the quantum affine symmetries and their restrictions are large
enough to characterize the S-matrices, they are likely to
characterize off-shell information as well.

\bigskip\bigskip
\noindent{\bf Acknowledgements}
\medskip
We would especially like to thank D. Bernard for many helpful
discussions in the course of this work.  We have also benefited
from discussions with P. Argyres, J. Cardy, T. Klassen, S. Mathur,
E. Melzer, F. Smirnov, and H. Tye.  We thank the organizers of
the Santa Barbara workshop on CFT for the opportunity to begin
our collaboration, and the organizers of RIMS-91 in Kyoto for
the occasion to present this work.  This work is supported in
part by the National Science Foundation.

\listrefs

\def\j#1#2#3#4{{
\left\{ \matrix{{1}&{#1}&{#2}\cr{1}&{#3}&{#4} \cr} \right\}_q
}}

\appendix{A}{Some q-6j symbols.}
\bigskip

\def\ta{{ \frac{3}{2} }}

Here we present explicit expressions for spin 1-1 q-6j symbols.  The
other q-6j symbols that are needed above can be found in \rBLii .
Below, $[n]$ denotes $[n]_q$, defined in \IIii .
$$\eqalignno{
\j jjjj &= \frac{ [2j] }{[2j+2] }
\( [2j+3][2j-1] - 1 \)
\cr
\noalign{\medskip}
\j {j+1}{j+1}{j}{j+1} &=
\j j{j+1}{j+1}{j+1}
= - \j {j+1}jjj \cr
&= -\j jj{j+1}j = \frac{[2]}{[2j+2]}
\cr
\noalign{\medskip}
\j j{j+1}jj &= \j jjj{j+1} = \frac{[2]}{[2j+2]}
\sqrt{ \frac{[2j+3]}{[2j+1]} }
\cr
\noalign{\medskip}
\j j{j-1}jj &= \j jjj{j-1} = - \frac{[2]}{[2j]}
\sqrt{ \frac{[2j-1]}{[2j+1]} }
\cr
\noalign{\medskip}
\j j{j+1}j{j+1} &= \j {j+\ha}{j-\ha}{j+\ha}{j-\ha} =
\frac{[2]}{[2j+2][2j+1] }
\cr
\noalign{\medskip}
\j j{j+1}{j+2}{j+1} &= \j j{j-1}{j-2}{j-1} = 1
\cr
\noalign{\medskip}
\j j{j+1}{j+1}j &= \j {j+1}jj{j+1} = \j jj{j+1}{j+1} \cr
&= \j {j+\ha}{j-\ha}{j+\ha}{j+\ta}
= \j {j+1}{j+1}jj  \cr
&= \j {j+\ha}{j+\ta}{j+\ha}{j-\ha}
= \inv{[2j+2]}
\sqrt{ [2j][2j+4] }
\cr}$$

\bye
\listrefs

\figures
\fig{1}{The contour of integration defining the adjoint action in
\IIxx .}
\fig{2}{The contour of integration in \seriii\ with endpoint $(x,t)$.}
\fig{3}{The product of braiding matrices representing the matrix
$M_{(j_{n+1} ,j_0 )}$.}

\bye